\documentclass[lettersize,journal]{IEEEtran}
\usepackage{amsmath,amsfonts}
\usepackage{algorithmic}
\usepackage{algorithm}
\usepackage{array}
\usepackage[caption=false,font=normalsize,labelfont=sf,textfont=sf]{subfig}
\usepackage{textcomp}
\usepackage{stfloats}
\usepackage{url}
\usepackage{verbatim}
\usepackage{booktabs}
\usepackage{multirow}
\usepackage{graphicx}
\usepackage[table,xcdraw]{xcolor}
\usepackage[normalem]{ulem}
\useunder{\uline}{\ul}{}
\usepackage{cite}
\usepackage{siunitx}
\DeclareSIUnit{\pixel}{pixel}
\usepackage{makecell}
\newcommand{\CI}[2]{{\scriptsize [#1, #2]}}
\begin{document}

\title{Multi-Beholder: Biomarker Prediction for Low-Grade Glioma with Multiple Instance Learning and One-Class Classification}
\author{Zijie Fang, Yihan Liu, Yifeng Wang, Xiangyang Zhang, Yang Chen, Changjing Cai, Yiyang Lin, Ying Han, Zhi Wang,~\IEEEmembership{Senior Member,~IEEE,} Shan Zeng, Jun Tan, Yongbing Zhang,~\IEEEmembership{Senior Member,~IEEE,} and Hong Shen
        % <-this % stops a space
\thanks{Zijie Fang and Yihan Liu contributed equally. Corresponding authors: Yongbing Zhang; Hong Shen.}
\thanks{Zijie Fang, Yang Chen, Yiyang Lin, and Zhi Wang are with Tsinghua Shenzhen International Graduate School, Tsinghua University, Shenzhen 518055, China }
\thanks{Yihan Liu, Xiangyang Zhang, Changjing Cai, Ying Han, and Shan Zeng are with the Department of Oncology, Xiangya Hospital, Central South University, Changsha 410008, China.}
\thanks{Yifeng Wang is with the School of Science, Harbin Institute of Technology (Shenzhen), Shenzhen 518055, China}
\thanks{Jun Tan is with the Department of Neurosurgery, Xiangya Hospital, Central South University, Changsha 410008, China}
\thanks{Yongbing Zhang is with 
the School of Computer Science and Technology, Harbin Institute of Technology (Shenzhen), Shenzhen 518055, China (email: ybzhang08@hit.edu.cn)}
\thanks{Hong Shen is with the Department of Oncology, Xiangya Hospital, Central South University, Changsha 410008, China. (email: hongshen2000@csu.edu.cn)}}
% \thanks{Manuscript received October 22, 2024.}}

% The paper headers
\markboth{Journal of \LaTeX\ Class Files,~Vol.~14, No.~8, August~2021}%
{Fang \MakeLowercase{\textit{et al.}}: Multi-Beholder}

\IEEEpubid{0000--0000/00\$00.00~\copyright~2021 IEEE}
% Remember, if you use this you must call \IEEEpubidadjcol in the second
% column for its text to clear the IEEEpubid mark.

\maketitle

\begin{abstract}
Biomarker detection is an indispensable part of the diagnosis and treatment of low-grade glioma (LGG). However, current LGG biomarker detection methods rely on expensive and complex molecular genetic testing, for which professionals are required to analyze the results, and intra-rater variability is often reported. To overcome these challenges, we propose an interpretable deep learning pipeline, named Multi-Biomarker Histomorphology Discoverer (Multi-Beholder), to predict the status of five biomarkers in LGG using only hematoxylin and eosin-stained whole slide images. Specifically, Multi-Beholder incorporates one-class classification into the multiple instance learning framework to achieve accurate instance-level pseudo-labeling, thereby complementing slide-level labels and improving prediction performance. Multi-Beholder demonstrates high performance on two LGG cohorts with diverse races and scanning protocols, with area under the receiver operating characteristic curve up to 0.973 on the internal-validated TCGA-LGG dataset and 0.820 on the external-validated Xiangya cohort. Moreover, the interpretability of Multi-Beholder allows for discovering quantitative and qualitative correlations between biomarker status and histomorphology characteristics. Our pipeline not only provides a novel approach for biomarker prediction, enhancing the applicability of molecular treatments for LGG patients but also facilitates the discovery of new mechanisms in molecular functionality and LGG progression. Code can be accessed at https://github.com/Vison307/Multi-Beholder.
\end{abstract}


\section{Introduction}
\label{intro}
\IEEEPARstart{G}{lioma} is one of the most prevalent primary brain tumours. According to the World Health Organization's grading system, gliomas can be divided into four grades \cite{weller2024glioma}, among which low-grade glioma (LGG) includes grades below III, and the median overall survival for grade II glioma is approximately 32.2 to 163.7 months \cite{ghaffari2020effect}. Although LGG is not as aggressive as high-grade glioma (HGG), it can still infiltrate normal tissue and may recur as an HGG \cite{houshyari2015comparative}. Therefore, early screening and treatment are vital for LGG patients. 

Presently, biomarker detection plays a significant role in LGG diagnosis and treatment. On the one hand, biomarker status can assist pathologists in subtyping. For instance, LGG cases with both isocitrate dehydrogenase (IDH1/2) mutation and combined loss of the short arm of chromosome 1 and the long arm of chromosome 19 (1p/19q codeletion) are diagnosed as oligodendrogliomas, which typically carry a favorable prognosis with longer progression-free survival \cite{yao2020human}. And the co-occurrence of alpha thalassemia/mental retardation syndrome X-linked (ATRX) mutation and IDH1/2 mutation suggests a higher likelihood of astrocytoma \cite{ohba2020correlation}. On the other hand, biomarker status helps plan treatments. Alkylating agent chemotherapy appears to be more effective in LGG patients with 1p/19q codeletion \cite{shaw2012randomized}, and the methylation status of the O6-Methylguanine-DNA-Methyltransferase (MGMT) promoter is a well-established predictive biomarker to determine the response to temozolomide \cite{everhard2006mgmt}. 

In the routine clinical workflow of LGG biomarker detection, pathologists rely on biopsy samples of the lesion tissues for molecular testing, which depends on wet laboratory techniques such as immunohistochemistry and DNA sequencing \cite{haase2018mutant}. These methods typically require additional reagents and specialized equipment, and interpreting the test results demands professional personnel. These factors considerably impede the promotion of biomarker detection, especially in resource-limited regions. Moreover, intra-rater inconsistency in the evaluation of detection results is frequently reported, resulting in resource wastage due to the requirement for re-testing. These issues in biomarker detection further hinder the improvement of overall survival in LGG patients.
\IEEEpubidadjcol

Since the U.S. Food and Drug Administration cleared whole-slide imaging (WSI) for primary pathology diagnosis in 2017, the digitization of glass slides has accelerated and is increasingly integrated into routine diagnostic workflows \cite{aggarwal2025artificial}. In parallel with the rapid growth of WSI repositories and advances in artificial intelligence, deep learning–based automated analysis has given rise to the emerging field of computational pathology. Because hematoxylin and eosin (H\&E)-stained histopathology sections are routinely acquired for diagnosing LGG, they provide a near-universal substrate for digitization into WSIs when slide scanning is available \cite{stone2024paediatric}. Consequently, deep learning models can leverage readily accessible WSIs to infer biomarker status, reducing reliance on additional molecular assays and supporting LGG diagnosis, risk stratification, and clinical decision-making \cite{chauhan2024ipd}.

\begin{figure}[!t]
    \centering
    \includegraphics[width=\linewidth]{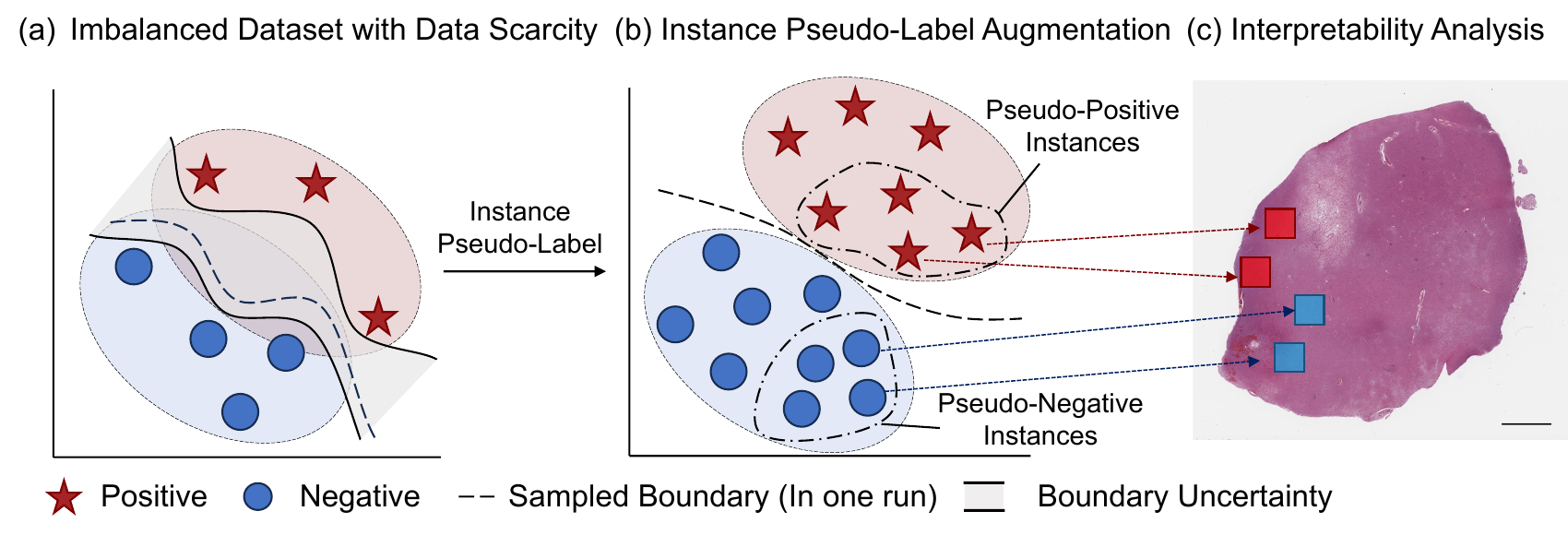}
    \caption{Illustration of the limitations of existing methods (a) and our proposed solution (b), which can be further used for both quantitative and qualitative interpretability analyses by identifying highly suspicious regions (c).}
    \label{fig:fig1.r1}
\end{figure}

However, WSIs usually consist of hundreds of millions of pixels to clearly reflect both local cell details and overall tissue morphology \cite{tu2023deep}. The huge size of WSIs brings tremendous challenges to train deep learning models due to memory limitations of graphic processing units (GPUs). To address this issue, some researchers adopt the multiple instance learning (MIL) framework into computational pathology, where WSIs are split into small patches, called instances, with the WSIs viewed as bags. In MIL, instance aggregators are designed to obtain bag features, enabling the training process of deep learning models based on bag-level labels \cite{wang2024advances}. Despite its effectiveness, MIL still suffers from the data-scarcity issue in computational pathology, especially for the biomarker prediction task in LGG, as shown in Fig.~\ref{fig:fig1.r1}(a). On one hand, due to the low incidence rate of LGG, the number of WSIs available for training MIL models is usually limited, making it difficult for the model to achieve sufficient training, which in turn leads to insufficient robustness of the learned feature representations and poorer separability of the feature space. On the other hand, due to significant differences in the occurrence probabilities of different gene mutations or methylation events, the LGG biomarker prediction task often suffers from severe label imbalance, making discriminative samples even scarcer. In this situation, model training is easily dominated by the majority class, making it difficult to fully learn the discriminative patterns of the minority class. Besides, data scarcity usually leads to significant fluctuations in the decision boundary under different data partitions or random initializations, reducing training stability and ultimately weakening the recognition ability and generalization performance for the minority class.

Compared with the limited number of WSIs, each huge WSI can be partitioned into an abundant number of instances. By assigning pseudo-positive and pseudo-negative labels to discriminative instances and incorporating them as auxiliary supervision for MIL training, we can effectively increase the amount of training evidence, enhance the separability of the learned feature space, and reduce decision-boundary uncertainty, thereby improving model stability and robustness, as illustrated in Fig.~\ref{fig:fig1.r1}(b). In addition, as shown in Fig.~\ref{fig:fig1.r1}(c), by mapping instance pseudo-labels back to the 2-dimensional spatial layout of the WSI, we can qualitatively visualize and quantitatively characterize the spatial distribution of biomarker-negative and biomarker-positive evidence (e.g., clustered regions, spatial heterogeneity, and the proportion of pseudo-positive or pseudo-negative instances). This provides intuitive and verifiable evidence to support model predictions, enabling clinically meaningful interpretability analysis and morphology-related discovery. However, obtaining instance-level annotations is usually time-consuming and labor-intensive. Thus, how to achieve effective instance-level supervision without manual labeling to further improve MIL performance remains an important challenge.

\begin{figure}[!t]
    \centering
    \includegraphics[width=\linewidth]{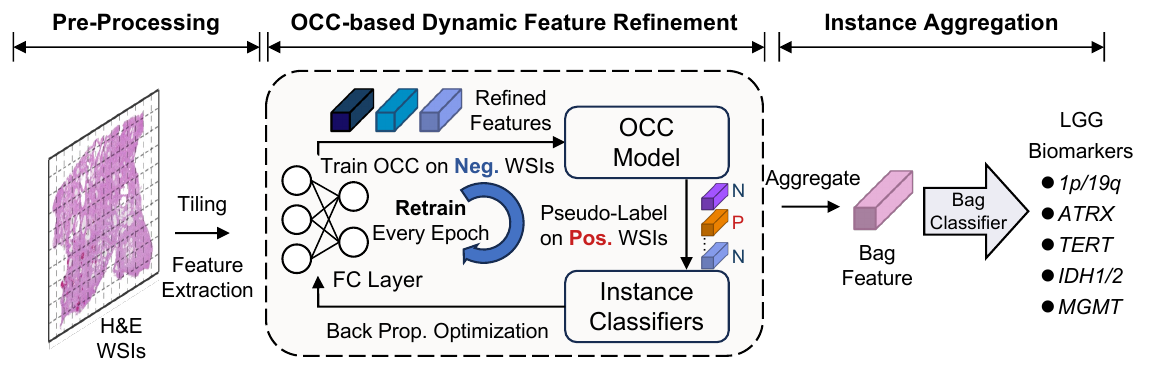}
    \caption{A concise schematic figure of the proposed Multi-Beholder pipeline. Neg., N: Negative; Pos., P: Positive; Back Prop.: Back propagation.}
    \label{fig:fig2.r1}
\end{figure}

To address this challenge, we propose a Multi-Biomarker Histomorphology Discoverer (Multi-Beholder) pipeline for LGG biomarker prediction. A concise schematic figure of the pipeline is shown in Fig.~\ref{fig:fig2.r1}. Specifically, after pre-processing H\&E-stained WSIs with instance feature extraction, we sufficiently utilize the true negative instances in negative WSIs to obtain a negative-positive decision boundary with a one-class classification (OCC) strategy. Then, the decision boundary of the OCC is employed to obtain the pseudo-labels of instances in positive WSIs. Finally, the pseudo-labels are exploited to refine the instance features by training instance-level classifiers. As MIL learns task-specific representations, its feature space becomes increasingly separable. Accordingly, we retrain the OCC model after each epoch to update instance pseudo-labels dynamically, which in turn refines MIL representations and improves bag-level biomarker prediction. The proposed Multi-Beholder successfully predicts the status of five biomarkers in LGG, including 1p/19q codeletion, ATRX mutation, IDH1/2 mutation, MGMT promoter methylation, and telomerase reverse transcriptase (TERT) promoter mutation, solely based on H\&E-stained WSIs over two different cohorts. Specifically, the main contributions of this paper are listed as follows:

\begin{itemize}
\item We present Multi-Beholder, a unified multi-biomarker prediction framework, for LGG that relies solely on routine H\&E-stained WSIs to predict the status of 1p/19q codeletion, IDH1/2 mutation, ATRX mutation, MGMT promoter methylation, and TERT promoter mutation.

\item We propose a dynamic, OCC-enabled instance pseudo-labeling strategy within Multi-Beholder. Specifically, an OCC model is iteratively trained using true negative instances mined from negative WSIs, and then used to infer pseudo-labels for instances in positive WSIs. These pseudo-labels are incorporated as auxiliary supervision to refine instance representations, thereby improving feature separability and robustness.

\item Experimental results on two cohorts demonstrate the effectiveness of Multi-Beholder. Furthermore, we conduct both quantitative and qualitative analyses with professional pathologists to investigate histomorphological correlates of biomarker status, laying the groundwork for future studies on biomarker-associated mechanisms.
\end{itemize} 
The paper is organized as follows. Section~\ref{intro} motivates LGG biomarker prediction from H\&E WSIs, outlines challenges of existing MIL methods, and summarizes the proposed OCC-based Multi-Beholder framework and contributions. Section~\ref{related works} reviews related work on MIL in computational pathology, deep learning-based biomarker prediction, and OCC in computer vision. Section~\ref{method} details the proposed method, and Section~\ref{results} reports comparative and ablation experiments along with interpretability analyses with 1p/19q codeletion as an exemplar. Section~\ref{conclusion} concludes the paper and discusses future directions. The \textbf{Supplementary Material} provides additional implementation details, extended interpretability and calibration results, decision curve analysis (DCA), further ablations, sensitivity analyses, and computational profiling to support reproducibility.

\section{Related Works}
\label{related works}
\subsection{MIL in Computational Pathology}
The giga-pixeled nature of WSIs makes it unfeasible to directly feed WSIs into deep learning models. To address this problem, MIL frameworks split WSIs into small patches, named instances, and develop aggregation methods to pool instances into bag-level representations. For example, Ilse et al. \cite{ilse2018attention} proposed an instance aggregation strategy based on the attention mechanism and designed the ABMIL framework. Based upon ABMIL, Lu et al. \cite{lu2021data} designed a clustering constraint to supervise the instance features in WSIs, improving the model's data efficiency and performance. To alleviate the problem of over-fitting resulting from the limited size of WSI datasets, Zhang et al. \cite{zhang2022dtfd} proposed a double-tier distillation MIL strategy named DTFD-MIL. By taking into consideration the dependencies between different instances, Li et al. \cite{li2021dual} proposed the DSMIL framework to model the relationships between the highest-scored instance and other instances to obtain a better instance aggregator. Shao et al. \cite{shao2021transmil} further utilized Transformer \cite{vaswani2017attention} to learn the interactions between every two instances and proposed the TransMIL framework. Zhao et al.~\cite{zhao2025ptcmil} introduced learnable prompt tokens to unify clustering and prediction in an end-to-end framework, enabling task-aware instance aggregation. However, the quadratic complexity of self-attention in Transformers often leads to substantial GPU memory demands and training time overhead, which limits scalability to gigapixel WSIs.

To mitigate this issue, recent studies have explored state space models (SSMs), a new sequence modeling paradigm with linear-time complexity, to build MIL frameworks. For example, Fillioux et al.~\cite{fillioux2023structured} leveraged a diagonal structured SSM for MIL and demonstrated its effectiveness on multi-class classification tasks. Yang et al.~\cite{yang2024mambamil} adopted selective structured state space models (Mamba) \cite{gu2024mamba} into MIL and proposed MambaMIL. Fang et al.~\cite{fang2024mammil} considered the non-uniform spatial distribution of patches in WSIs by representing a WSI as an undirected graph and proposing a spatial-aware Mamba scanning mechanism. However, all the previous studies neglected the property that the instances in negative WSIs are almost negative. We fully exploit this property to assign pseudo-labels to the instances for feature refinement based on an OCC strategy, which largely improves the performance of LGG biomarker prediction.

\subsection{Deep Learning in Biomarker Prediction}
In recent years, deep learning has been widely adopted in computer-aided healthcare, including biomarker prediction, due to its objectivity and strong empirical performance. For example, Schirris et al.~\cite{schirris2022deepsmile} proposed the DeepSMILE framework and showed that genome-related labels such as microsatellite instability and homologous recombination deficiency can be predicted by only H\&E-stained pathological images on the TCGA colorectal cancer and breast cancer cohorts. Shamai et al.~\cite{shamai2022deep} evaluated the feasibility of predicting PD-L1 status in breast cancer from H\&E images using convolutional neural networks. Zhang et al.~\cite{zhang2024exploring} integrated CT imaging with clinicopathological data and developed a deep learning-based radiomics model, termed Deep-RadScore, to predict gene mutations such as EGFR, KRAS, and TP53 in non-small cell lung cancer patients.

However, studies focusing on LGG remain relatively limited, mainly due to restricted data availability and the intrinsic difficulty arising from the relatively homogeneous histomorphology of LGG tissues. Some works have attempted to infer molecular status in glioma patients using MRI~\cite{shboul2020prediction}. In addition, Zhang et al.~\cite{zhang2025machine} developed a machine learning model based on radiomic features extracted from contrast-enhanced MRI to predict Interleukin-18 status. Nevertheless, compared with histopathological images, MRI has limited spatial resolution and therefore cannot provide cellular- and microstructure-level morphological evidence. Consequently, several studies have explored the prediction of specific biomarkers under limited histopathology data, such as IDH1/2 mutation, p53 mutation, and MGMT promoter methylation status from H\&E images~\cite{jiang2021predicting,liu2020isocitrate,li2023vision}. However, existing approaches generally lack systematic investigations into model interpretability, and the large number of parameters in deep learning models further increases the difficulty of understanding and explaining their decision mechanisms. Enhancing interpretability not only helps patients and clinicians build trust in model decisions, but also facilitates the investigation of tumor progression mechanisms and biomarker functionality. Motivated by these considerations, we conduct a comprehensive interpretability study of Multi-Beholder to identify potential associations between biomarker status and histomorphological patterns.

\subsection{OCC in Computer Vision}
\begin{figure*}[!t]
    \centering
\includegraphics[width=0.7\linewidth]{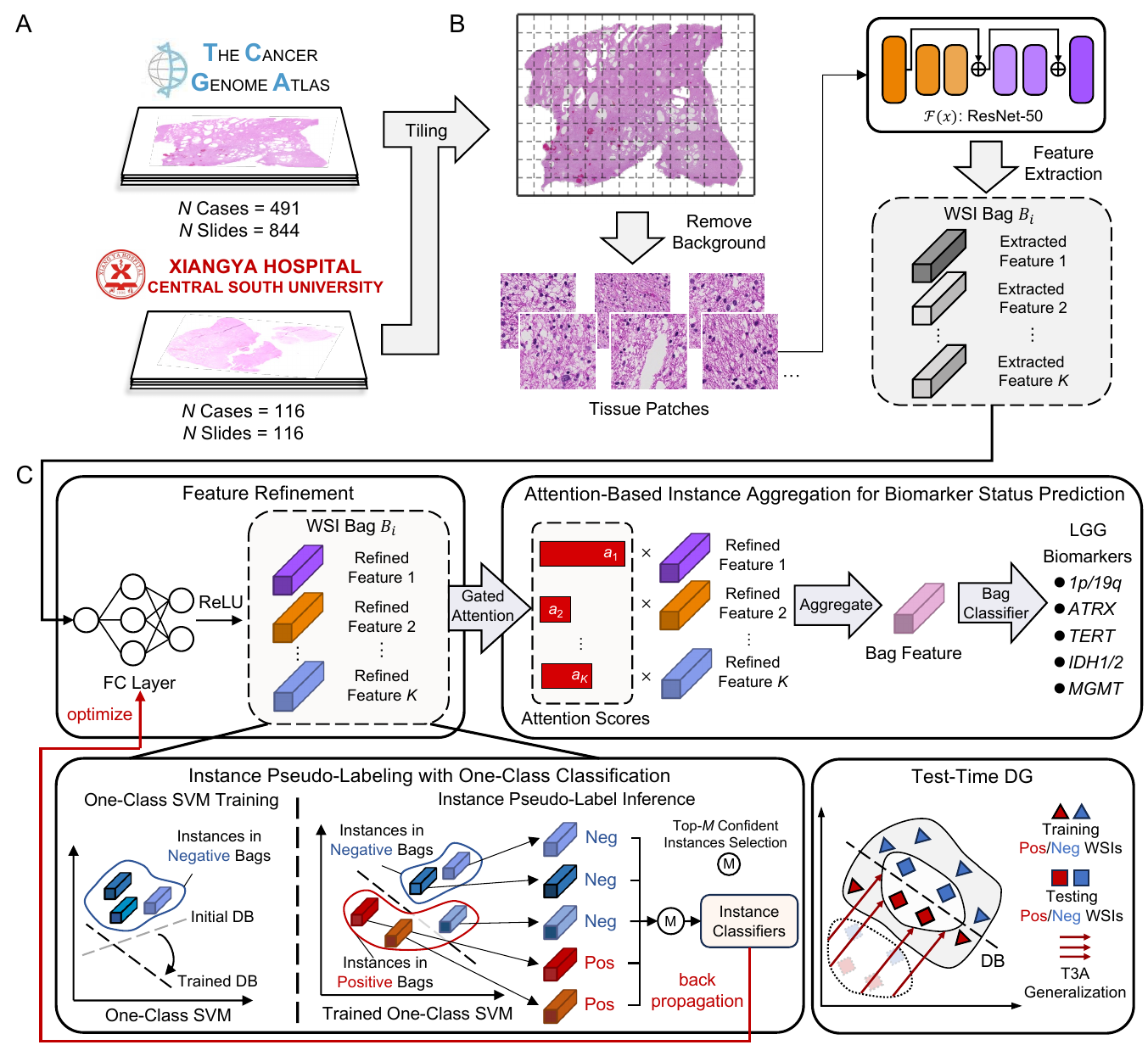}
    \caption{An overview of the Multi-Beholder pipeline. (a) Data sources of the Multi-Beholder pipeline. (b) The pre-processing of the Multi-Beholder pipeline. (c) The workflow of biomarker prediction. SVM: support vector machine; Neg: negative; Pos: positive; DB: decision boundary; DG: domain generalization.}
    \label{fig:1}
\end{figure*}

OCC only requires data in a single category during the model training process and can determine whether the test samples belong to the same category as the training samples. Therefore, OCC is widely used in computer vision tasks such as anomaly detection and novelty detection \cite{dhar2023doc,wang2020hierarchical}. In computational pathology, recent studies have explored OCC strategies to learn negative distributions from biomarker-negative or normal samples. For example, Xu et al.~\cite{xu2024novelty} introduced a deep OCC constraint into MIL and proposed a compact negative instance feature mining strategy, which encourages instance embeddings from negative bags to form a more compact distribution in the latent space, leading to improved PD-L1 status prediction in non-small cell lung cancer. However, this work primarily focuses on a single biomarker application and does not systematically examine whether a unified framework can support the prediction of multiple tumor biomarkers. Dang et al.~\cite{dang2025abnormality} trained an abnormality detector on normal WSIs using a Gaussian mixture variational autoencoder, and leveraged the trained model to extract informative patches from WSIs, thereby reducing training computational costs. By further integrating multimodal features, their method supports cancer diagnosis and subtype classification. Nevertheless, this study mainly uses the OCC-style model as an instance selection module, rather than exploiting its potential to assign instance-level pseudo-labels for explicitly enhancing feature-space separability in MIL. Additionally, Nejat et al.~\cite{nejat2024creating} constructed a normal tissue atlas from normal WSIs and applied OCC for anomaly detection to filter patches in tumor WSIs, aiming to increase the representativeness of retained patches and reduce the computational burden of the downstream retrieval task. Nevertheless, this study mainly emphasizes improving the efficiency of WSI retrieval and does not further investigate the potential of OCC for biomarker prediction. Motivated by these observations, we develop a unified MIL framework that performs dynamic OCC-based instance pseudo-labeling and conduct systematic external validation on independent cohorts, enabling robust multi-biomarker prediction and pathology-grounded interpretability analyses.

\section{Method}
\label{method}
This paper proposes a pipeline named Multi-Beholder to predict the status of LGG biomarkers. An overview of Multi-Beholder is illustrated in Fig. \ref{fig:1}.
Since each WSI contains hundreds of millions of pixels, we first cut the WSIs into non-overlapping equal-sized patches. Next, considering the labels of the patches are unavailable, we adopt the MIL framework where each WSI is viewed as a bag and the patches in the WSI are regarded as instances. Finally, the features of all the instances are aggregated to a bag feature, enabling the supervision of the pipeline with bag-level labels. Besides, considering that the instances in negative bags are almost negative, we introduce the OCC strategy into the MIL framework. The OCC strategy can fully exploit the true negative instances in negative WSIs to obtain a reliable negative-positive instance decision boundary for instance pseudo-labeling. Compared with bag-level labels, the abundant instance pseudo-labels can enhance the discriminability of the instance features, ultimately improving the performance of biomarker prediction. To further avoid the performance degradation caused by differences in WSI preparation process, such as scanning and staining procedures, we apply the test-time template augmentation (T3A) \cite{iwasawa2021test} strategy to achieve test-time domain generalization. To improve readability, the key symbols and hyperparameters along with their definitions are shown in Table \ref{tab:symbols}. 
\begin{table}[t]
\centering
{
\caption{Definitions of key symbols and hyperparameters.}
\label{tab:symbols}
\begin{tabular}{p{0.12\linewidth} p{0.75\linewidth}}
\toprule
Symbol & Definition \\
\midrule
$B_i$ & A WSI (bag) represented as a set of patch instances. \\
$x_{ij}$ & The $j$-th patch instance sampled from a WSI $B_i$. \\
$h_{ij}$ & Extracted instance feature of $x_{ij}$ (before refinement). \\
$z_{ij}$ & Refined instance feature produced by the refinement layer. \\
$a_{ij}$ & Instance attention weight used for aggregation. \\
$Z_i$ & Aggregated bag (WSI) feature. \\
$\hat{y}_{ij}$, $\hat{Y}_i$ & Instance ($\hat{y}_{ij}$) or bag ($\hat{Y}_i$) predictions from MIL. \\
$\tilde{Y}_i$ & Final bag prediction after T3A. \\
$\bar{d}_{ij}$ & Anomaly score used to rank informative instances. \\
$\mathbf{w}$,$\rho$ & OCSVM normal vector and bias defining the decision boundary. \\
$\mathbb{S}$ & Support sets in T3A. \\
$\nu$ & OCSVM hyperparameter controlling the fraction of support vectors and outliers. \\
$M$ & Number of confident instances selected in each WSI for pseudo-labeling. \\
$r$ & Ratio of top anomaly-score instances excluded as noisy samples in a positive WSI. \\
$\alpha_1$, $\alpha_2$ & Weights balancing bag-level and instance-level losses; $\alpha_1+\alpha_2=1$. \\
\bottomrule
\end{tabular}
}
\end{table}

\subsection{Review of the MIL framework}
The giga-pixeled nature of WSIs makes it necessary to cut the WSIs into patches, whose labels cannot be obtained. Therefore, the biomarker prediction problem is a typical weakly-supervised binary classification task, which can be solved under the MIL framework. The MIL framework considers WSIs as bags, denoted as $\{B_i\}$, where $i$ stands for the index of a bag. Each bag is composed of $K$ different instances denoted by $\{x_{i1},x_{i2},\cdots,{x_{iK}}\}$ ($K$ may vary across different WSIs). We only have access to the bag-level label $Y_i$ but cannot obtain the labels $\{y_{i1},y_{i2},\cdots,y_{iK}\}$ of the instances. 

The basic assumption of MIL is that if the label of a bag is positive, there is at least one instance in the bag whose label is positive as well. Otherwise, if the label of a bag is negative, all instances in the bag should have negative labels, i.e.,
\begin{equation}
    Y_i=1 \Leftrightarrow \exists y_{i j}=1, j=1, \cdots K \Leftrightarrow \max _j\left\{y_{i j}\right\}=1,
\end{equation}
\begin{equation}
\label{eq2}
    Y_i=0 \Leftrightarrow \forall y_{i j}=0, j=1, \cdots, K \Leftrightarrow \max _j\left\{y_{i j}\right\}=0.
\end{equation}

This study considers codeletion, mutation, or methylation of a gene or a promoter as a positive label. Since each WSI corresponds to a specific biomarker status, they form a sample point $\{B_i,Y_i\}$, where $Y_i\in\{0,1\}$.

\subsection{Pre-processing and Instance Feature Extraction}
To significantly reduce the GPU memory occupation, we feed the instance features rather than patches into the MIL framework. Specifically, we first feed each patch $x_{i j} \in \mathbb{R}^{256 \times 256 \times 3}$ in each WSI into a ResNet50 \cite{he2016deep} pre-trained on ImageNet \cite{deng2009imagenet} for instance feature extraction by
\begin{equation}
h_{i j}=\mathcal{F}\left(x_{i j}\right),
\end{equation}
where $h_{i j} \in \mathbb{R}^{1024}$ is the extracted feature, and $\mathcal{F}$ is the ResNet50. After feature extraction, each WSI can be represented as a collection of feature vectors, i.e., $B_i=\left\{h_{i 1}, h_{i 2}, \cdots, h_{i K}\right\}$.

\subsection{OCC-based Dynamic Feature Refinement}
In the instance feature extraction procedure, the utilized ResNet50 is pre-trained on natural images. However, the difference between natural and pathological images may lead to unsatisfactory feature extraction. To address this problem, we place a trainable instance feature refinement layer after ResNet50, which is composed of a fully connected (FC) layer with a ReLU activation calculated by
\begin{equation}
z_{i j}=\operatorname{ReLU}\left(\mathbf{W}^{\mathrm{T}} h_{i j}+\mathbf{b}\right),
\end{equation}
where $\mathbf{W} \in \mathbb{R}^{1024 \times 512}$ and $\mathbf{b} \in \mathbb{R}^{512}$ are the weight and bias of the FC layer, respectively. $\operatorname{ReLU}(x)=\max(0,x)$ stands for the ReLU function, and $z_{i j} \in \mathbb{R}^{512}$ is the refined feature vector.

However, directly supervising the instance feature refinement layer requires accurate labels of the instances, which are not available under the MIL framework. To address this problem, we adopt the OCC strategy to make full use of the true negative instances in negative WSIs to obtain an accurate negative-positive instance decision boundary, thereby enabling the discrimination of negative and positive instances in the positive WSIs to assign precise instance pseudo-labels. Specifically, one-class support vector machine (OCSVM) \cite{scholkopf1999support} is employed in Multi-Beholder. Let $\{z_1^{-}, z_2^{-}, \cdots, z_{N^{-}}^{-}\}$ be the set of refined instance features in the negative WSIs, where $N^-$ represents the total number of instances in all negative WSIs. The basic idea of OCSVM is to find a hyperplane farthest to the zero point in the feature space, which ensures all the negative training instances are on the same side of the normal vector of the hyperplane. In testing, the negative instances are located on the side of the normal vector of the hyperplane, while the positive instances are located on the other side. The objective function of OCSVM is defined as
\begin{equation}
\label{occ objective}
\min _{\mathbf{w}, \rho, \xi_i} \frac{1}{2}\|\mathbf{w}\|^2-\rho+\frac{1}{v N^{-}} \sum_{i=1}^{N^{-}} \xi_i,
\end{equation}
s.t.
\begin{equation}
\left\langle\mathbf{w}, z_i^{-}\right\rangle \geq \rho-\xi_i, \quad 1 \leq i \leq N^{-},
\end{equation}
\begin{equation}
\label{occ c2}
\xi_i \geq 0,1 \leq i \leq N^{-}.
\end{equation}
Here, $\mathbf{w}$ and $\rho$ are trainable parameters of the OCSVM, representing the normal vector and the intercept of the hyperplane, respectively. $\xi_i$ is the slack variable. $\nu \in(0,1]$ is a hyper-parameter used to determine an upper bound on the training error and a lower bound on the support vector ratio.

After training OCSVM utilizing (\ref{occ objective})-(\ref{occ c2}), we can judge whether an instance in a positive WSI belongs to the negative category or not. Let $B_i^+$ be a testing positive bag composed of instances $\left\{z_{i 1}^{+}, z_{i 2}^{+}, \cdots, z_{i K^{+}}^{+}\right\}$, where $K^+$ is the number of instances in the bag. The directed distance of each instance $z_{ij}^+$ to the hyperplane $\mathbf{w}$ without considering the intercept is calculated by
\begin{equation}
d_{i j}=\left\langle\mathbf{w}, z_{i j}^{+}\right\rangle .
\end{equation}
The larger the directed distance, the higher the likelihood of the instance being negative. We define the opposite value $\bar{d}_{i j}=-d_{i j}$ as the anomaly score of each instance in positive bags. Eventually, we assign pseudo-labels to the instances in positive bags based on the anomaly scores. To ensure the accuracy of the pseudo-labels, we only select the instances with the highest or lowest anomaly scores as confident instances for pseudo-labeling. Specifically, all instances are first sorted according to the anomaly score $\bar{d}_{i j}$. Then, the $M$ instances with the minimum anomaly scores are labeled as negative, and those with the $M$ maximum anomaly scores are annotated as positive. It is worth noting that some noisy patches, such as pen marks, shadows, and bubbles in the WSIs, are totally different from the tissue patches and have extremely high anomaly scores. Hence, we do not select these patches with the highest anomaly score in $r$ percentage, where $r$ is a hyperparameter.

Finally, the confident instances along with their pseudo-labels are used as additional supervision to train the instance feature refinement layer. To take full advantage of the instance-level supervision, we design two separate instance classifiers for both negative and positive instances. The negative instance classifier classifies negative and non-negative instances and vice versa. Specifically, the two classifiers are built by
\begin{equation}
\hat{y}_{i j}^{-}=\operatorname{Softmax}\left(\mathbf{W}_{-}^{\mathrm{T}} z_{i j}+\mathbf{b}_{-}\right),
\end{equation}
\begin{equation}
\hat{y}_{i j}^{+}=\operatorname{Softmax}\left(\mathbf{W}_{+}^{\mathrm{T}} z_{i j}+\mathbf{b}_{+}\right),
\end{equation}
where $\mathbf{W}_{-} \in \mathbb{R}^{512 \times 2}$ and $\mathbf{W}_{+} \in \mathbb{R}^{512 \times 2}$ are learnable weights, $\mathbf{b}_{-} \in \mathbb{R}^2$ and $\mathbf{b}_{+} \in \mathbb{R}^2$ are learnable biases, and $\hat{y}_{i j}^{-}$ and $\hat{y}_{i j}^{+}$ represent the probability of $z_{ij}$ being a negative instance or a positive instance, respectively.

Given that the OCSVM takes as input the refined instance features produced by the refinement layer, and that the corresponding embedding space progressively learns task-specific representations and becomes increasingly separable during MIL training, the OCC decision boundary should be updated accordingly. To achieve this, we retrain the OCSVM from scratch at the beginning of each epoch using the latest refined instance features extracted from negative WSIs, enabling the OCSVM to dynamically adapt to the evolving embedding space. Based on the dynamically updated OCSVM, we assign pseudo-labels to instances in positive WSIs in an epoch-wise manner, thereby improving pseudo-label reliability and providing sustained positive guidance for subsequent feature refinement and MIL optimization.

Considering that we only assign pseudo-labels to the confident instances in positive WSIs, to ensure data balance, we only assign negative labels to the top $M$ and bottom $M$ instances for negative WSIs as well by measuring the attention score calculated in the instance aggregation process, which is described in the following subsection.

\subsection{Attention-Based Instance Aggregation}
To predict the slide-level biomarker status, it is necessary to aggregate the feature representations of all instances in the bag and obtain a bag-level feature. Traditional instance aggregation operators in MIL include average pooling, max pooling, and others. However, these operators without trainable parameters have poor generalization ability. To address this issue, we adopt an attention-based feature aggregation method \cite{ilse2018attention}. Technically, for each refined instance feature $z_{ij}$, denote its attention score to the bag-level feature as $a_{ij}$. The final bag-level representation is aggregated by the weighted sum of all instance features over the attention scores by
\begin{equation}
Z_i=\sum_{j=1}^K a_{i j} \cdot z_{i j},
\end{equation}
where $Z_i \in \mathbb{R}^{512}$ is the aggregated bag-level feature. In this work, the attention score $a_{ij}$ is obtained by the gated attention mechanism as
\begin{equation}
a_{i j}=\frac{\exp \left\{\mathbf{w}_{\text{A}}^{\mathrm{T}}\left(\tanh \left(\mathbf{V}_{\text{A}}^{\mathrm{T}} z_{i j}\right) \odot \operatorname{sigmoid}\left(\mathbf{U}_{\text{A}}^{\mathrm{T}} z_{i j}\right)\right)\right\}}{\sum_{j=1}^K \exp \left\{\mathbf{w}_{\text{A}}^{\mathrm{T}}\left(\tanh \left(\mathbf{V}_{\text{A}}^{\mathrm{T}} z_{i j}\right) \odot \operatorname{sigmoid}\left(\mathbf{U}_{\text{A}}^{\mathrm{T}} z_{i j}\right)\right)\right\}},
\end{equation}
where $\mathbf{w}_{\text{A}} \in \mathbb{R}^{D \times 1}$, $\mathbf{V}_{\text{A}} \in \mathbb{R}^{512 \times D}$, and $\mathbf{U}_{\text{A}} \in \mathbb{R}^{512 \times D}$ are learnable weights. $D$ is a manually determined hyper-parameter controlling the dimension of the latent features.

After obtaining the bag-level representations, an FC layer activated by the Softmax function is utilized as the bag classifier to obtain the predicted probabilities of the biomarker status as
\begin{equation}
\hat{Y}_i=\operatorname{Softmax}\left(\mathbf{W}_{\mathrm{BAG}}^{\mathrm{T}} Z_i+\mathbf{b}_{\mathrm{BAG}}\right),
\end{equation}
where $\mathbf{W}_{\mathrm{BAG}} \in \mathbb{R}^{512 \times 2}$ and $\mathbf{b}_{\mathrm{BAG}} \in \mathbb{R}^2$ are the learnable weight and bias, respectively.

\subsection{Optimization Objectives}
We use cross entropy as the loss function for negative and positive instance classifiers. For an instance with pseudo-label $y_{ij}$, the loss of the negative instance classifier is expressed as
\begin{equation}
L^{-}\left(z_{i j}\right)=-\left[\left(1-y_{i j}\right) \log \left(\hat{y}_{i j}^{-}[1]\right)+y_{i j} \log \left(\hat{y}_{i j}^{-}[0]\right)\right],
\end{equation}
where $\hat{y}_{i j}^{-}[0]$ and $\hat{y}_{i j}^{-}[1]$, i.e., the first and second component of $\hat{y}_{i j}^{-}$, represent the predicted non-negative and negative probability from the negative instance classifier.

Similarly, the loss of the positive instance classifier can be expressed as
\begin{equation}
L^{+}\left(z_{i j}\right)=-\left[y_{i j} \log \left(\hat{y}_{i j}^{+}[1]\right)+\left(1-y_{i j}\right) \log \left(\hat{y}_{i j}^{+}[0]\right)\right] .
\end{equation}

The bag-level classifier also uses the cross-entropy loss, which is defined as
\begin{equation}
L^{\mathrm{BAG}}\left(B_i\right)=-\left[Y_i \log \left(\hat{Y}_i[1]\right)+\left(1-Y_i\right) \log \left(\hat{Y}_i[0]\right)\right].
\end{equation}

For each WSI $B_i$, the total loss function is defined as the weighted sum of the losses of the instance classifiers and the bag-level classifier, expressed as
\begin{equation}
L\left(B_i\right)=\alpha_1 L^{\mathrm{BAG}}\left(B_i\right)+\frac{\alpha_2\sum_{z_{i j} \in C\left(B_i\right)} L^{-}\left(z_{i j}\right)+L^{+}\left(z_{i j}\right)}{2\left|C\left(B_i\right)\right|} .
\end{equation}
Here $C(B_i)$ represents the features of the confident instances in $B_i$. $|\cdot|$ represents the cardinality of a set. $\alpha_1$ and $\alpha_2$ are manually set weights, which satisfy $\alpha_1+\alpha_2=1$.

\subsection{Test-Time Domain Generalization}
Due to distinctions in ethnicity, staining, scanners, etc., there are often differences between WSIs used in training and testing, leading to a substantial performance drop when a model trained on the training set is directly applied to the test set. To solve this problem, Multi-Beholder employs T3A into the MIL framework.
Specifically, two support sets, $\mathbb{S}^{(0)}$ and $\mathbb{S}^{(1)}$, are maintained for negative and positive WSIs, respectively, with the initial element being the weight of the bag-level classifier corresponding to the biomarker status. Namely, for the weight of the trained bag-level classifier $\mathbf{W}_{\mathrm{BAG}}$, we have
\begin{equation}
\mathbb{S}^{(i)}=\{\mathbf{W}_{\mathrm{BAG}}[:, i]\} \in \mathbb{R}^{512}, i=0,1.
\end{equation}

Next, for each test WSI $B_i$, we first feed it into the trained Multi-Beholder and get a raw prediction $\hat{Y}_i$. Then, we normalize the bag representation $Z_i$ and add it to the support set corresponding to $\hat{Y}_i$, i.e.,
\begin{equation}
\mathbb{S}^{(\hat{Y}_i)} \leftarrow \mathbb{S}^{(\hat{Y}_i)} \cup\left\{\frac{Z_i}{\|Z_i\|}\right\},
\end{equation}
where $\|\cdot\| $ represents the L-2 norm.

Finally, the average of each support set is calculated as the new weight for the bag classifier, and the actually predicted label $\tilde{Y}_i$ for bag $B_i$ with bag feature $Z_i$ is calculated by
\begin{equation}
    \tilde{Y}_i=\operatorname{argmax}_j\left\{\frac{\exp \left(<c_j, Z_i>\right)}{\exp \left(<c_0, Z_i>\right)+\exp \left(<c_1, Z_i>\right)}\right\},
\end{equation}
where
\begin{equation}
\label{eq21}
c_j=\frac{1}{\left|\mathbb{S}^{(j)}\right|} \sum_{z \in \mathbb{S}^{(j)}} z, j=0,1.
\end{equation}

In practice, considering $\hat{Y}_i$ may be wrong, only features with raw prediction entropy below the $C$-th largest among the support set are considered in (\ref{eq21}), where $C$ is a hyperparameter.

\section{Experimental Results}
\label{results}
\subsection{Data Sources and Pre-Processing}
A public LGG dataset from The Cancer Genome Atlas (TCGA) program, TCGA-LGG, and an in-house dataset collected from the Xiangya Hospital are utilized in the experiments for evaluation. Specifically, the Xiangya dataset is utilized as an external validation of the Multi-Beholder trained on TCGA-LGG to evaluate the generalization ability of the pipeline. This study was reviewed and approved by the Xiangya Hospital Medical Ethics Committee of Central South University (No. 202012235) in December 2020. Statistics of the two datasets are listed in \textbf{Supplementary Table XVII}.

\subsubsection{Public TCGA-LGG Dataset}
We collected 844 WSIs from 491 cases diagnosed with LGG in TCGA to build the TCGA-LGG dataset. It is worth noting that the TCGA-LGG dataset exhibits substantial class imbalance, which significantly increases the complexity of the prediction task. As shown in \textbf{Supplementary Table XVII}, IDH1/2 mutation and MGMT promoter methylation are highly skewed toward the positive class, with over 80\% of cases labeled as mutated or methylated, respectively. In contrast, biomarkers such as 1p/19q codeletion and ATRX mutation display the opposite trend, with lower proportions of positive cases. TERT promoter mutation presents a moderate imbalance favoring the wild-type class, compounded by a considerable proportion of missing labels. These distributional disparities introduce challenges in model training, particularly in accurately identifying minority-class patterns. Therefore, training and evaluating models on the TCGA-LGG dataset provides a rigorous test of their robustness to class imbalance and demonstrates that the proposed Multi-Beholder framework maintains strong predictive performance under these adverse conditions.

In TCGA-LGG, all WSIs are scanned in the SVS format. In the dataset, 379 cases only have one WSI, while a few cases have multiple WSIs, with a maximum of 17 WSIs per case. For cases with multiple WSIs, we assume that the biomarker labels of all WSIs are consistent with the biomarker status of the case. For each WSI, we follow the CLAM pre-processing pipeline \cite{lu2021data} to segment background regions and tile the tissue into $256\times256$ pixel patches at $20\times$ magnification (approximately \SI{0.5}{\micro\metre\per\pixel}). We discard a WSI if its tissue content is insufficient, i.e., it contains no continuous tissue region larger than $4096\times4096$ pixels at $40\times$ magnification (approximately \SI{0.25}{\micro\metre\per\pixel}). After pre-processing, we have 831 WSIs from 485 cases. Then, considering that not all cases have the status of all five biomarkers, we group the WSIs with a specific biomarker status into a dataset. For each biomarker, we use a 10-fold Monte Carlo cross-validation to divide the dataset into training, validation, and test sets under a ratio of 8:1:1. WSIs from the same case are ensured in the same set.

\subsubsection{In-house Xiangya Dataset}
We retrospectively collected WSIs from patients diagnosed with LGG from the Xiangya Hospital, Central South University. The Xiangya dataset contains 116 WSIs from 116 cases, where 94 WSIs are in the SDPC format, and 22 WSIs are in the QPTIFF format. Only 1p/19q codeletion, ATRX mutation, IDH1/2 mutation, and MGMT promoter methylation status are recorded. Besides, the ATRX mutation and IDH1/2 mutation status of some cases are marked as indeterminate (neither positive nor negative), and we discard these ambiguous labels. We pre-process these WSIs as those in TCGA-LGG. 
\begin{table*}[!t]
{
\caption{Comparison results on TCGA-LGG for each biomarker and overall performance. Values are reported as mean with 95\% CIs in brackets. \textbf{Bold} and \underline{underlined} entries denote the best and second-best results, respectively. $^{*}$ indicates AUROC $p<0.05$ when comparing Multi-Beholder with the corresponding method. $N$ denotes the total number of available WSIs; $N^{+}$ and $N^{-}$ denote the numbers of positive and negative WSIs, respectively, with percentages reported in parentheses.}
\label{tab:main}
\resizebox{\linewidth}{!}{%
\begin{tabular}{@{}l|ccc|l|ccc@{}}
\toprule
\multicolumn{1}{c|}{}                                  & Accuracy                            & Recall                              & AUROC                               & \multicolumn{1}{c|}{}                                  & Accuray                             & Recall                              & AUROC                               \\ \cmidrule(lr){2-4} \cmidrule(l){6-8} 
\multicolumn{1}{c|}{\multirow{-2}{*}{\textbf{Method}}} & \multicolumn{3}{c|}{\cellcolor[HTML]{EFEFEF}1p/19q ($N$=831; $N^+$=325 (39.11\%); $N^-$=506 (69.89\%)}                                                             & \multicolumn{1}{c|}{\multirow{-2}{*}{\textbf{Method}}} & \multicolumn{3}{c}{\cellcolor[HTML]{EFEFEF}ATRX ($N$=828; $N^+$=285 (34.42\%); $N^-$=543 (65.58\%)}                                                                \\ \midrule
ABMIL$^{*}$                                            & 0.764 \CI{0.709}{0.805}             & 0.732 \CI{0.671}{0.780}             & 0.830 \CI{0.786}{0.871}             & ABMIL$^{*}$                                            & 0.730 \CI{0.685}{0.768}             & \underline{0.687 \CI{0.635}{0.733}} & 0.812 \CI{0.781}{0.840}             \\
CLAMSB$^{*}$                                          & 0.730 \CI{0.682}{0.779}             & 0.708 \CI{0.647}{0.762}             & 0.812 \CI{0.771}{0.851}             & CLAMSB$^{*}$                                          & 0.725 \CI{0.676}{0.771}             & 0.680 \CI{0.626}{0.731}             & 0.798 \CI{0.759}{0.831}             \\
CLAMMB$^{*}$                                          & 0.736 \CI{0.688}{0.785}             & 0.716 \CI{0.669}{0.764}             & 0.839 \CI{0.805}{0.871}             & CLAMMB$^{*}$                                          & 0.716 \CI{0.661}{0.763}             & 0.665 \CI{0.606}{0.722}             & 0.790 \CI{0.731}{0.840}             \\
DSMIL$^{*}$                                            & 0.708 \CI{0.653}{0.757}             & 0.673 \CI{0.611}{0.731}             & 0.804 \CI{0.761}{0.840}             & DSMIL$^{*}$                                            & 0.695 \CI{0.640}{0.752}             & 0.621 \CI{0.557}{0.693}             & 0.755 \CI{0.702}{0.808}             \\
TransMIL$^{*}$                                         & 0.607 \CI{0.529}{0.679}             & 0.599 \CI{0.532}{0.671}             & 0.678 \CI{0.586}{0.767}             & TransMIL$^{*}$                                         & 0.642 \CI{0.588}{0.692}             & 0.512 \CI{0.477}{0.556}             & 0.570 \CI{0.502}{0.652}             \\
DTFD-MIL$^{*}$                                         & 0.740 \CI{0.680}{0.797}             & 0.719 \CI{0.654}{0.775}             & 0.838 \CI{0.799}{0.873}             & DTFD-MIL$^{*}$                                         & \underline{0.743 \CI{0.696}{0.786}} & \textbf{0.699 \CI{0.658}{0.737}}    & \underline{0.818 \CI{0.783}{0.847}} \\
MambaMIL$^{*}$                                         & 0.690 \CI{0.613}{0.752}             & 0.672 \CI{0.617}{0.725}             & 0.823 \CI{0.775}{0.861}             & MambaMIL$^{*}$                                         & 0.665 \CI{0.596}{0.726}             & 0.585 \CI{0.526}{0.655}             & 0.758 \CI{0.702}{0.810}             \\
S4Model$^{*}$                                          & 0.732 \CI{0.687}{0.774}             & 0.711 \CI{0.665}{0.757}             & 0.789 \CI{0.741}{0.832}             & S4Model$^{*}$                                          & 0.658 \CI{0.613}{0.702}             & 0.592 \CI{0.559}{0.626}             & 0.720 \CI{0.667}{0.769}             \\
PTC-MIL$^{*}$                                          & \underline{0.775 \CI{0.739}{0.810}} & \textbf{0.756 \CI{0.716}{0.797}}    & \underline{0.854 \CI{0.813}{0.892}} & PTC-MIL$^{*}$                                          & 0.688 \CI{0.632}{0.746}             & 0.645 \CI{0.586}{0.708}             & 0.756 \CI{0.702}{0.807}             \\ \midrule
Multi-Beholder (ours)                                  & \textbf{0.787 \CI{0.707}{0.858}}    & \underline{0.751 \CI{0.664}{0.828}} & \textbf{0.940 \CI{0.876}{0.980}}    & Multi-Beholder (ours)                                  & \textbf{0.752 \CI{0.679}{0.822}}    & 0.652 \CI{0.575}{0.733}             & \textbf{0.973 \CI{0.951}{0.991}}    \\ \midrule
\textbf{}                                              & \multicolumn{3}{c|}{\cellcolor[HTML]{EFEFEF}IDH1/2 ($N$=828; $N^+$=702 (84.78\%); $N^-$=126 (15.22\%)}                                                             & \textbf{}                                              & \multicolumn{3}{c}{\cellcolor[HTML]{EFEFEF}MGMT ($N$=831; $N^+$=700 (84.24\%); $N^-$=131 (15.76\%)}                                                                \\ \midrule
ABMIL                                                  & 0.852 \CI{0.824}{0.878}             & 0.626 \CI{0.597}{0.655}             & 0.817 \CI{0.785}{0.852}             & ABMIL$^{*}$                                            & 0.838 \CI{0.803}{0.868}             & 0.507 \CI{0.490}{0.529}             & 0.575 \CI{0.492}{0.646}             \\
CLAMSB                                                & 0.853 \CI{0.825}{0.876}             & \textbf{0.674 \CI{0.624}{0.723}}    & \textbf{0.831 \CI{0.793}{0.865}}    & CLAMSB$^{*}$                                          & 0.840 \CI{0.798}{0.875}             & 0.504 \CI{0.496}{0.515}             & 0.615 \CI{0.564}{0.676}             \\
CLAMMB                                                & \textbf{0.869 \CI{0.853}{0.885}}    & 0.625 \CI{0.567}{0.694}             & 0.824 \CI{0.790}{0.858}             & CLAMMB$^{*}$                                          & \textbf{0.855 \CI{0.827}{0.878}}    & \underline{0.535 \CI{0.496}{0.590}} & \underline{0.644 \CI{0.565}{0.727}} \\
DSMIL$^{*}$                                            & 0.856 \CI{0.836}{0.876}             & 0.521 \CI{0.496}{0.547}             & 0.739 \CI{0.678}{0.791}             & DSMIL$^{*}$                                            & \textbf{0.855 \CI{0.831}{0.877}}    & 0.500 \CI{0.500}{0.500}             & 0.584 \CI{0.510}{0.645}             \\
TransMIL$^{*}$                                         & \underline{0.861 \CI{0.845}{0.878}} & 0.500 \CI{0.500}{0.500}             & 0.526 \CI{0.427}{0.619}             & TransMIL$^{*}$                                         & \textbf{0.855 \CI{0.831}{0.877}}    & 0.500 \CI{0.500}{0.500}             & 0.425 \CI{0.346}{0.499}             \\
DTFD-MIL                                               & 0.848 \CI{0.829}{0.868}             & 0.645 \CI{0.577}{0.717}             & 0.782 \CI{0.718}{0.836}             & DTFD-MIL$^{*}$                                         & 0.850 \CI{0.822}{0.875}             & 0.531 \CI{0.492}{0.597}             & 0.614 \CI{0.552}{0.679}             \\
MambaMIL$^{*}$                                         & \underline{0.861 \CI{0.845}{0.878}} & 0.500 \CI{0.500}{0.500}             & 0.672 \CI{0.611}{0.718}             & MambaMIL$^{*}$                                         & \textbf{0.855 \CI{0.831}{0.877}}    & 0.500 \CI{0.500}{0.500}             & 0.498 \CI{0.429}{0.577}             \\
S4Model$^{*}$                                          & 0.809 \CI{0.752}{0.857}             & 0.577 \CI{0.513}{0.654}             & 0.750 \CI{0.702}{0.799}             & S4Model$^{*}$                                          & 0.829 \CI{0.798}{0.858}             & 0.493 \CI{0.469}{0.515}             & 0.560 \CI{0.497}{0.635}             \\
PTC-MIL$^{*}$                                          & 0.854 \CI{0.830}{0.877}             & \underline{0.646 \CI{0.615}{0.680}} & 0.752 \CI{0.708}{0.792}             & PTC-MIL$^{*}$                                          & \underline{0.852 \CI{0.828}{0.874}} & 0.509 \CI{0.499}{0.527}             & 0.538 \CI{0.443}{0.621}             \\ \midrule
Multi-Beholder (ours)                                  & 0.849 \CI{0.820}{0.876}             & 0.617 \CI{0.565}{0.670}             & \underline{0.825 \CI{0.798}{0.855}} & Multi-Beholder (ours)                                  & 0.844 \CI{0.806}{0.877}             & \textbf{0.541 \CI{0.495}{0.612}}    & \textbf{0.650 \CI{0.583}{0.712}}    \\ \midrule
\multicolumn{1}{c|}{\textbf{}}                         & \multicolumn{3}{c|}{\cellcolor[HTML]{EFEFEF}TERT ($N$=489; $N^+$=242 (49.49\%); $N^-$=247 (50.51\%)}                                                               & \multicolumn{1}{c|}{\textbf{}}                         & \multicolumn{3}{c}{\cellcolor[HTML]{EFEFEF}Overall}                                                             \\ \midrule
ABMIL$^{*}$                                            & 0.706 \CI{0.625}{0.776}             & 0.711 \CI{0.644}{0.774}             & 0.812 \CI{0.755}{0.866}             & ABMIL$^{*}$                                            & \underline{0.778 \CI{0.753}{0.800}} & 0.653 \CI{0.639}{0.666}             & 0.769 \CI{0.746}{0.791}             \\
CLAMSB                                                & \underline{0.724 \CI{0.684}{0.765}} & \underline{0.722 \CI{0.680}{0.762}} & \underline{0.827 \CI{0.769}{0.881}} & CLAMSB$^{*}$                                          & 0.775 \CI{0.753}{0.798}             & 0.657 \CI{0.632}{0.682}             & \underline{0.777 \CI{0.759}{0.794}} \\
CLAMMB$^{*}$                                          & 0.697 \CI{0.615}{0.769}             & 0.693 \CI{0.627}{0.755}             & 0.791 \CI{0.727}{0.854}             & CLAMMB$^{*}$                                          & 0.774 \CI{0.750}{0.796}             & 0.647 \CI{0.625}{0.668}             & \underline{0.777 \CI{0.762}{0.794}} \\
DSMIL$^{*}$                                            & 0.684 \CI{0.627}{0.740}             & 0.688 \CI{0.639}{0.738}             & 0.784 \CI{0.737}{0.832}             & DSMIL$^{*}$                                            & 0.759 \CI{0.734}{0.784}             & 0.600 \CI{0.578}{0.623}             & 0.733 \CI{0.703}{0.761}             \\
TransMIL$^{*}$                                         & 0.587 \CI{0.515}{0.670}             & 0.563 \CI{0.500}{0.647}             & 0.640 \CI{0.534}{0.750}             & TransMIL$^{*}$                                         & 0.711 \CI{0.686}{0.737}             & 0.535 \CI{0.511}{0.563}             & 0.568 \CI{0.517}{0.620}             \\
DTFD-MIL$^{*}$                                         & 0.683 \CI{0.616}{0.733}             & 0.697 \CI{0.643}{0.744}             & 0.821 \CI{0.770}{0.869}             & DTFD-MIL$^{*}$                                         & 0.773 \CI{0.748}{0.796}             & \underline{0.658 \CI{0.632}{0.684}} & 0.775 \CI{0.753}{0.797}             \\
MambaMIL$^{*}$                                         & 0.576 \CI{0.515}{0.639}             & 0.591 \CI{0.533}{0.650}             & 0.727 \CI{0.655}{0.790}             & MambaMIL$^{*}$                                         & 0.729 \CI{0.698}{0.763}             & 0.570 \CI{0.543}{0.598}             & 0.695 \CI{0.673}{0.719}             \\
S4Model$^{*}$                                          & 0.668 \CI{0.614}{0.720}             & 0.669 \CI{0.621}{0.715}             & 0.748 \CI{0.693}{0.797}             & S4Model$^{*}$                                          & 0.739 \CI{0.714}{0.765}             & 0.608 \CI{0.584}{0.637}             & 0.713 \CI{0.689}{0.742}             \\
PTC-MIL$^{*}$                                          & 0.704 \CI{0.658}{0.749}             & 0.706 \CI{0.663}{0.748}             & 0.784 \CI{0.730}{0.836}             & PTC-MIL$^{*}$                                          & 0.775 \CI{0.750}{0.795}             & 0.653 \CI{0.639}{0.666}             & 0.737 \CI{0.701}{0.770}             \\
\midrule
Multi-Beholder (ours)                                  & \textbf{0.727 \CI{0.640}{0.802}}    & \textbf{0.736 \CI{0.661}{0.804}}    & \textbf{0.835 \CI{0.775}{0.887}}    & Multi-Beholder (ours)                                  & \textbf{0.792 \CI{0.748}{0.829}}    & \textbf{0.659 \CI{0.624}{0.690}}    & \textbf{0.845 \CI{0.823}{0.864}}    \\ \bottomrule
\end{tabular}%
}
}
\end{table*}
\subsection{Experimental Environment and Settings}
The proposed Multi-Beholder is trained on an NVIDIA RTX 3090 GPU (32GB) with a batch size of 1. The latent dimension for attention computation is set to $D{=}128$. We use Adam~\cite{kingma2014adam} with a learning rate of $5\times10^{-4}$, $(\beta_1,\beta_2){=}(0.9,0.999)$, and a weight decay of $10^{-4}$. The training process of each fold takes approximately 0.5 hours with around 520 MB GPU memory, and the average inference time of each fold takes around 3 seconds with less than 300 MB memory with pre-extracted instance features. We use ResNet50 \cite{he2016deep} as the default instance feature extractor. The ablation study on different feature extractors is described in \textbf{Supplementary Section VII}. It is worth noting that retraining the OCSVM after each epoch only adds minimal overhead (less than 10 minutes per fold). Overall, the computational cost is modest, and the proposed Multi-Beholder can be applied to large-scale datasets, indicating good scalability. More details about the computation profile are in \textbf{Supplementary Section II}.

In the comparative experiments, we use accuracy, recall, and area under the receiver operating characteristic curve (AUROC) as main metrics. We also report precision, F1-score, and area under the precision-recall curve (also known as average precision, AP) in \textbf{Supplementary Section IX}. All metrics are macro-averaged except for accuracy, which is micro-averaged. Early stopping is applied based on the validation AUROC with patience of 10 epochs and maximum of 200 epochs, and the checkpoint with the best validation AUROC is used for testing. Detailed implementation and hyperparameter settings are provided in \textbf{Supplementary Section I}.

\subsection{Statistical Analysis}
We perform a 10-fold Monte Carlo cross-validation with case-wise splitting, ensuring that all WSIs from the same case are kept in the same split. We report the mean test performance across folds and estimate 95\% CIs via nonparametric bootstrap over folds with 10,000 resamples, where folds are treated as the resampling unit. For AUROC comparisons, we run DeLong’s test on the WSI-level predictions within each fold’s held-out test set and combine the resulting $p$-values across folds using Fisher’s method. Overall AUROC is further compared using a paired Wilcoxon signed-rank test across folds, where each fold contributes a single value obtained by averaging its metrics over all biomarkers.

\subsection{Interpretability Analyses}
Since the proposed Multi-Beholder additionally determines the positive and negative biomarker status on the instance level, both qualitative and quantitative analyses can be realized to discover the underlying correlations between histomorphology characteristics and biomarkers, which makes one of the biggest contributions of Multi-Beholder. An overview of the interpretability analysis process is shown in \textbf{Supplementary Fig. 22}, and is described in detail in this subsection.

\subsubsection{Visualization of Attention Maps}
For attention map visualization, we choose the trained Multi-Beholder with the best test AUROC and randomly select one WSI from the test set. We first calculate the raw attention score of each patch $z_{ij}$ by $\mathbf{w}_{\text{A}}^{\mathrm{T}}(\tanh (\mathbf{V}_{\text{A}}^{\mathrm{T}} z_{i j})\odot \operatorname{sigmoid}(\mathbf{U}_{\text{A}}^{\mathrm{T}} z_{i j}))$. Then, we scale the scores by Min-Max normalization. Finally, instances whose normalized attention scores are among the largest 10\% are visualized by applying a heatmap upon the WSI.

\subsubsection{Visualization of Instance Distribution Maps}
For the visualization of instance distribution maps, we first reduce the refined instance features to two dimensions using t-SNE \cite{van2008visualizing}. Then, we uniformly select up to 2,000 instances for plotting. To draw the decision boundary of the OCC on the 2-dimensional plot, we utilize support vector regression (SVR) \cite{drucker1996support} to estimate the decision boundary by feeding the dimension-reduced instance features and the instance labels from the trained OCSVM into the SVR. For ATRX mutation and TERT promoter mutation, because there are too few or even no predicted negative instances, it is difficult to train the SVR. So, we do not plot the estimated decision boundary. The attention scores are normalized by Min-Max normalization among the selected instances in the instance distribution maps. 

\subsubsection{Cell Type Proportion Calculation with HoverNet}
To calculate the proportions of different cell types in the top 10\% attended patches, we utilize the HoverNet \cite{graham2019hover} pre-trained on the PanNuke \cite{gamper2019pannuke,gamper2020pannuke} dataset, which is composed of H\&E-stained patches under different magnifications from 19 different tissues, including the brain tumor, with over 200,000 precise-labeled nuclei annotations. The PanNuke dataset is officially split into three folds. We train the HoverNet using the first fold and validate it on the second fold, with the third fold being the test set. Finally, the HoverNet with the best validation loss is selected for cell type proportion calculation.

\subsection{Comparison Results}
\begin{table*}[t]
\centering
{
\caption{The performance of directly transferring Multi-Beholder trained on TCGA-LGG to the Xiangya cohort without retraining. Values are reported as mean with 95\% CIs. WSI-level class distributions are 1p/19q ($N{=}63$, $N^{+}{=}19$ (30.16\%), $N^{-}{=}44$ (69.84\%)); ATRX ($N{=}109$, $N^{+}{=}33$ (30.28\%), $N^{-}{=}76$ (69.72\%)); IDH1/2 ($N{=}73$, $N^{+}{=}44$ (60.27\%), $N^{-}{=}29$ (39.73\%)); and MGMT ($N{=}64$, $N^{+}{=}26$ (40.62\%), $N^{-}{=}38$ (59.38\%)).}
\label{tab:5}
\renewcommand{\arraystretch}{1.15}
\setlength{\tabcolsep}{4pt}
\resizebox{\linewidth}{!}{
\begin{tabular}{lcccccc}
\toprule
Biomarker & Accuracy & Recall & Precision & F1-Score & AUROC & AP \\
\midrule
1p/19q  & 0.760 \CI{0.737}{0.783} & 0.650 \CI{0.592}{0.709} & 0.707 \CI{0.619}{0.763} & 0.640 \CI{0.567}{0.707} & 0.820 \CI{0.798}{0.836} & 0.629 \CI{0.604}{0.656} \\
ATRX    & 0.606 \CI{0.566}{0.647} & 0.593 \CI{0.575}{0.610} & 0.591 \CI{0.570}{0.614} & 0.568 \CI{0.544}{0.593} & 0.653 \CI{0.640}{0.667} & 0.471 \CI{0.443}{0.500} \\
IDH1/2  & 0.662 \CI{0.640}{0.685} & 0.599 \CI{0.562}{0.637} & 0.734 \CI{0.690}{0.774} & 0.559 \CI{0.500}{0.617} & 0.708 \CI{0.684}{0.733} & 0.778 \CI{0.751}{0.805} \\
MGMT    & 0.456 \CI{0.417}{0.502} & 0.530 \CI{0.504}{0.559} & 0.389 \CI{0.274}{0.505} & 0.383 \CI{0.317}{0.459} & 0.647 \CI{0.601}{0.685} & 0.593 \CI{0.543}{0.641} \\
\bottomrule
\end{tabular}%
}
}
\end{table*}

To validate the effectiveness of the proposed pipeline, we compare Multi-Beholder with eight state-of-the-art methods, ABMIL \cite{ilse2018attention}, CLAM \cite{lu2021data}, DSMIL \cite{li2021dual}, TransMIL \cite{shao2021transmil}, DTFD-MIL \cite{zhang2022dtfd}, MambaMIL \cite{yang2024mambamil}, S4Model \cite{fillioux2023structured}, and PTC-MIL \cite{zhao2025ptcmil}. It is worth noting that CLAM has two implementations with single or multiple attention branches, and we denote CLAM with a single branch as CLAMSB and that with multiple branches as CLAMMB. The comparison results are shown in Table~\ref{tab:main} and \textbf{Supplementary Table~XI--XVI}. The ROC curves and precision--recall (PR) curves are also illustrated in \textbf{Supplementary Fig.~17--21} for better reference.

Results show that Multi-Beholder successfully classifies patients with and without 1p/19q codeletion, achieving an AUROC of 0.940 on the test sets. \textbf{Supplementary Table~XI} shows that Multi-Beholder achieves the best precision (0.811) and AP (0.897) as well. From the curves in \textbf{Supplementary Fig.~17}, Multi-Beholder consistently maintains a higher true-positive rate at low false-positive rates on the ROC curve, and it also preserves higher precision over a wide range of recall on the PR curve, indicating a better overall performance.

We then apply the Multi-Beholder pipeline for ATRX mutation prediction. Results in Table~\ref{tab:main} show that Multi-Beholder achieves an AUROC of 0.973, outperforming all the comparison methods. Multi-Beholder also achieves the best accuracy under a 0.5 classification threshold. As shown in \textbf{Supplementary Table~XII}, Multi-Beholder further achieves the best precision and AP. From the ROC and PR curves in \textbf{Supplementary Fig.~18}, it can be observed that the ROC curve of Multi-Beholder stays above other methods, and its PR curve shows a clear precision advantage across most recall levels.

Next, the performance on IDH1/2 mutation prediction is evaluated. Table~\ref{tab:main} shows that Multi-Beholder achieves an AUROC of 0.825 and detailed results in \textbf{Supplementary Table~XIII} show that Multi-Beholder achieves an AP of 0.961, which is the best among all comparison methods. The ROC and PR curves are shown in \textbf{Supplementary Fig.~19}. It can be seen that although the ROC and PR curves of several strong methods are close to each other, the proposed Multi-Beholder still belongs to the first tier of performance.

We further evaluate MGMT promoter methylation prediction on TCGA-LGG. Results in Table~\ref{tab:main} demonstrate that Multi-Beholder achieves the best AUROC. It is worth noting that, although Multi-Beholder does not obtain the best accuracy, some best-performed methods predict all test WSIs into the same category under the fixed threshold. For example, DSMIL, TransMIL, and MambaMIL achieve the highest accuracy of 0.855, but their recalls are all 0.500, indicating that these comparison methods classify all samples into the same category. Therefore, a higher accuracy cannot imply better performance. However, the proposed Multi-Beholder does not suffer from this issue and achieves the best recall among all methods. Moreover, \textbf{Supplementary Table~XIV} shows that Multi-Beholder also achieves the best F1-score and AP, while maintaining second-best precision. From \textbf{Supplementary Fig.~20}, ROC curves of different methods are relatively close on this task, while Multi-Beholder shows a consistent advantage. The PR curve also indicates that Multi-Beholder maintains competitive precision at lower recall levels.

Besides, Table~\ref{tab:main} shows that Multi-Beholder achieves the top on all main metrics for TERT promoter mutation prediction. \textbf{Supplementary Table~XV} also illustrates that Multi-Beholder achieves the best AP and the second-best precision and F1-score compared with other methods. The ROC and PR curves shown in \textbf{Supplementary Fig.~21} reveal that Multi-Beholder has consistently favorable ROC and PR trends compared with competing baselines as well.

Finally, we calculate the overall performance of each method by averaging the metrics over all the five biomarkers. As shown in Table~\ref{tab:main} and \textbf{Supplementary Table~XVI}, Multi-Beholder achieves the best overall accuracy, recall, precision, AUROC, and AP among all methods. Although Multi-Beholder does not achieve the best overall F1-score, it is very close to the best-performing method, indicating that Multi-Beholder provides robust and consistent overall performance across multiple biomarker prediction tasks on TCGA-LGG.

\subsection{Generalization Ability of Multi-Beholder}
\label{sec:f}
To better assess the generalization ability of Multi-Beholder, we directly transfer the model trained on TCGA-LGG to our in-house Xiangya cohort, which differs from TCGA-LGG in staining protocols and patient ethnicities. The external validation results in Table~\ref{tab:5} demonstrate that Multi-Beholder retains encouraging cross-center generalization. Without any retraining, Multi-Beholder achieves an AUROC of 0.820, an accuracy of 0.760, and a precision of 0.707 for 1p/19q codeletion prediction, suggesting a balanced error profile under domain shift. Similar generalization trends are observed for IDH1/2 mutation prediction. Notably, IDH1/2 attains a relatively strong AP of 0.778, together with a high precision of 0.734.
However, for ATRX mutation prediction, the Multi-Beholder pipeline attains relatively modest performance. This may be attributed to the more global morphological effect of ATRX on tissue appearance. As shown in \textbf{Supplementary Fig.~13(b)}, the OCC module tends to assign pseudo-positive labels to all patches in ATRX-mutant WSIs, reducing the discriminativeness of selected pseudo-labeled instances and making it difficult to form a well-separated decision boundary. The domain shift between TCGA-LGG and Xiangya may further amplify this issue and degrade prediction performance.

For MGMT promoter methylation prediction, directly transferring the model to Xiangya yields a mean AUROC of 0.647, which is close to the in-domain AUROC of 0.650 on TCGA-LGG, indicating that the ranking ability is largely preserved. However, the accuracy is relatively low (around 0.45), suggesting that MGMT-related morphological cues are subtle and challenging under cross-center shift. This observation is consistent with the comparative results on TCGA-LGG, where most methods achieve AUROCs around 0.60 and tend to collapse to predicting a single class (recall $\approx$ 0.5). Moreover, our quantitative interpretability analysis in \textbf{Supplementary Fig.~15(b)} shows that only a very small fraction of tissue regions (on average 0.79\%) in MGMT-methylated WSIs are identified as pseudo-positive instances, further supporting that MGMT-associated evidence is extremely sparse and thus difficult for OCC to reliably extract sufficient, morphologically separable positive evidence from positive WSIs.

\subsection{Ablation Study on Model Performance With versus Without OCC}
To validate the reasonability of exploiting OCC, an ablation study is conducted to compare the performance with and without the OCC strategy on the TCGA-LGG dataset. For comparison fairness, we do not use T3A. Table \ref{tab:6} shows that incorporating OCC consistently improves performance for most biomarkers. Specifically, AUROC increases from 0.827 to 0.838 (+0.011) for 1p/19q, 0.815 to 0.825 (+0.010) for IDH1/2, 0.636 to 0.650 (+0.014) for MGMT, and 0.822 to 0.835 (+0.013) for TERT. AP also improves for 1p/19q (+0.019), MGMT (+0.010), TERT (+0.014), and IDH1/2 (+0.004). While OCC yields a slight performance drop for ATRX, this may be because ATRX-related morphology is more global (\textbf{Supplementary Fig. 13}), causing true positive instances within positive WSIs to be mislabeled as negative. Overall, the quantitative gains support the effectiveness of integrating OCC into the MIL framework.

\begin{table*}[!t]
\centering
{
\caption{The performance of Multi-Beholder on TCGA-LGG with and without the OCC strategy. Values are reported as mean with 95\% CIs in brackets. \textbf{Bold} entries denote the best results. $^{*}$ indicates AUROC $p<0.05$ when comparing Multi-Beholder with the corresponding variant without OCC. WSI-level class distributions are 1p/19q ($N{=}831$, $N^{+}{=}325$ (39.11\%), $N^{-}{=}506$ (60.89\%)); ATRX ($N{=}828$, $N^{+}{=}285$ (34.42\%), $N^{-}{=}543$ (65.58\%)); IDH1/2 ($N{=}828$, $N^{+}{=}702$ (84.78\%), $N^{-}{=}126$ (15.22\%)); MGMT ($N{=}831$, $N^{+}{=}700$ (84.24\%), $N^{-}{=}131$ (15.76\%)); and TERT ($N{=}489$, $N^{+}{=}242$ (49.49\%), $N^{-}{=}247$ (50.51\%)).}
\label{tab:6}
\resizebox{\linewidth}{!}{%
\begin{tabular}{llllllll}
\toprule
Biomarker & OCC & Accuracy & Recall & Precision & F1 & AUROC & AP \\
\midrule

\multirow{2}{*}{1p/19q} 
& $\times^\ast$ & 0.750 \CI{0.707}{0.791} & 0.728 \CI{0.672}{0.783} & \textbf{0.770 \CI{0.742}{0.799}} & 0.714 \CI{0.652}{0.772} & 0.827 \CI{0.771}{0.881} & 0.767 \CI{0.706}{0.830} \\
& $\checkmark$ & \textbf{0.759 \CI{0.699}{0.811}} & \textbf{0.743 \CI{0.671}{0.804}} & 0.736 \CI{0.625}{0.808} & \textbf{0.722 \CI{0.624}{0.798}} & \textbf{0.838 \CI{0.795}{0.875}} & \textbf{0.786 \CI{0.731}{0.833}} \\
\midrule

\multirow{2}{*}{ATRX}
& $\times$ & \textbf{0.759 \CI{0.713}{0.801}} & \textbf{0.711 \CI{0.642}{0.780}} & \textbf{0.763 \CI{0.711}{0.811}} & \textbf{0.699 \CI{0.632}{0.765}} & \textbf{0.819 \CI{0.776}{0.859}} & \textbf{0.683 \CI{0.631}{0.737}} \\
& $\checkmark$ & 0.729 \CI{0.671}{0.782} & 0.657 \CI{0.595}{0.721} & 0.718 \CI{0.655}{0.777} & 0.646 \CI{0.570}{0.723} & 0.817 \CI{0.781}{0.850} & 0.671 \CI{0.620}{0.722} \\
\midrule

\multirow{2}{*}{IDH1/2}
& $\times$ & \textbf{0.852 \CI{0.828}{0.878}} & 0.605 \CI{0.571}{0.640} & \textbf{0.701 \CI{0.654}{0.750}} & \textbf{0.614 \CI{0.577}{0.653}} & 0.815 \CI{0.779}{0.856} & 0.957 \CI{0.943}{0.971} \\
& $\checkmark$ & 0.849 \CI{0.820}{0.876} & \textbf{0.617 \CI{0.565}{0.670}} & 0.691 \CI{0.584}{0.800} & 0.605 \CI{0.549}{0.661} & \textbf{0.825 \CI{0.798}{0.855}} & \textbf{0.961 \CI{0.948}{0.973}} \\
\midrule

\multirow{2}{*}{MGMT}
& $\times^\ast$ & \textbf{0.844 \CI{0.814}{0.872}} & 0.519 \CI{0.496}{0.552} & 0.472 \CI{0.420}{0.538} & 0.486 \CI{0.455}{0.525} & 0.636 \CI{0.585}{0.693} & 0.904 \CI{0.889}{0.921} \\
& $\checkmark$ & \textbf{0.844 \CI{0.806}{0.877}} & \textbf{0.541 \CI{0.495}{0.612}} & \textbf{0.517 \CI{0.443}{0.610}} & \textbf{0.521 \CI{0.466}{0.603}} & \textbf{0.650 \CI{0.583}{0.712}} & \textbf{0.914 \CI{0.892}{0.931}} \\
\midrule

\multirow{2}{*}{TERT}
& $\times^\ast$ & 0.713 \CI{0.663}{0.764} & 0.716 \CI{0.670}{0.762} & \textbf{0.730 \CI{0.683}{0.776}} & 0.706 \CI{0.654}{0.758} & 0.822 \CI{0.773}{0.869} & 0.829 \CI{0.772}{0.884} \\
& $\checkmark$ & \textbf{0.727 \CI{0.640}{0.802}} & \textbf{0.736 \CI{0.661}{0.804}} & 0.721 \CI{0.586}{0.816} & \textbf{0.708 \CI{0.595}{0.797}} & \textbf{0.835 \CI{0.775}{0.887}} & \textbf{0.843 \CI{0.780}{0.896}} \\
\bottomrule
\end{tabular}%
} % end resizebox
}
\end{table*}

\subsection{Interpretability Analysis Results}
In this subsection, we describe the discovered histomorphology characteristics correlated with 1p/19q codeletion by using the proposed pipeline. The interpretability analyses of other four biomarkers are stated in \textbf{Supplementary Section VIII}.

\subsubsection{Attention Map Analyses}
\begin{figure}[!t]
    \centering
    \includegraphics[width=0.9\linewidth]{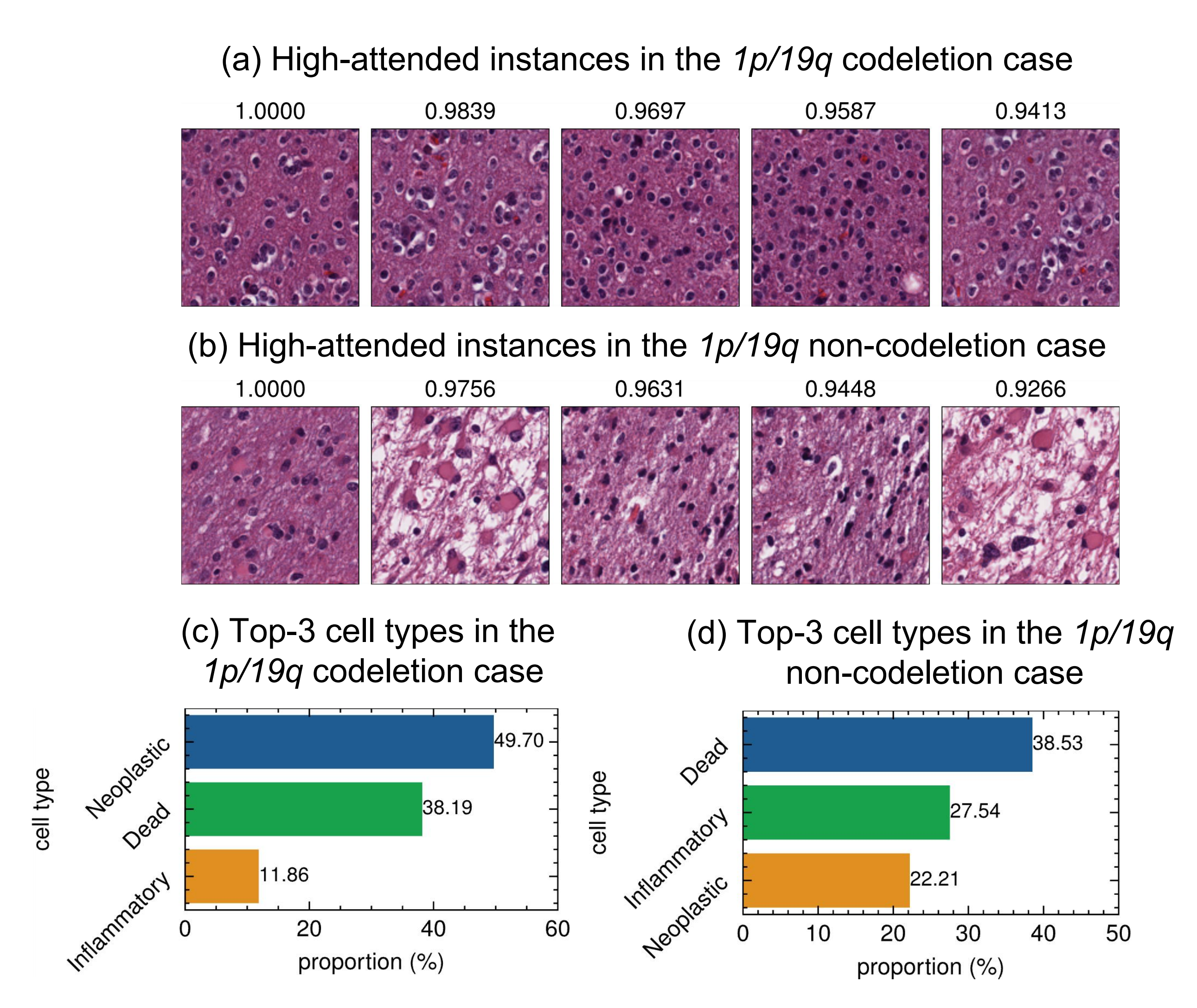}
    \caption{The enlarged patches and cell type proportions of the high-attended regions in a 1p/19q codeletion and non-codeletion case. The value above each patch represents the normalized attention score.}
    \label{fig:2}
\end{figure}

The Multi-Beholder discovers the correlations between 1p/19q codeletion and histomorphology characteristics. As shown in \textbf{Supplementary Fig. 23(a)}, attention visualization on a 1p/19q codeletion case illustrates that dense glial cells can be observed by darker staining in the attended regions, indicating that 1p/19q codeletion may lead to improved glial cell density. The enlarged top 5 attended regions in Fig. \ref{fig:2}(a) show that the perinuclear halo frequently appears. In comparison, as shown in Fig. \ref{fig:2}(b) and \textbf{Supplementary Fig. 23(b)}, the attention map of a case without 1p/19q codeletion implies that glial cells are sparse with severe atypia. Besides, eosinophilic granular bodies and Rosenthal fibers can be observed, suggesting that they may indicate a lower probability of 1p/19q codeletion. To quantitatively reveal the effect of 1p/19q codeletion at a cell level, we further count the proportion of diverse cell types in the top 10\% attended patches. Results in Fig. \ref{fig:2}(c) show that the majority of the attended cells are neoplastic in the 1p/19q codeletion case, with dead cells being the second majority. In comparison, in the 1p/19q non-codeletion case, the dead cells appear the most frequently, with the second being inflammatory cells, which indicates more inflammatory reactions in LGG patients without 1p/19q codeletion, as is shown in Fig. \ref{fig:2}(d). 

\subsubsection{Visualization of Instance Distribution and Label Proportion}
\begin{figure}[!t]
    \centering
    \includegraphics[width=0.7\linewidth]{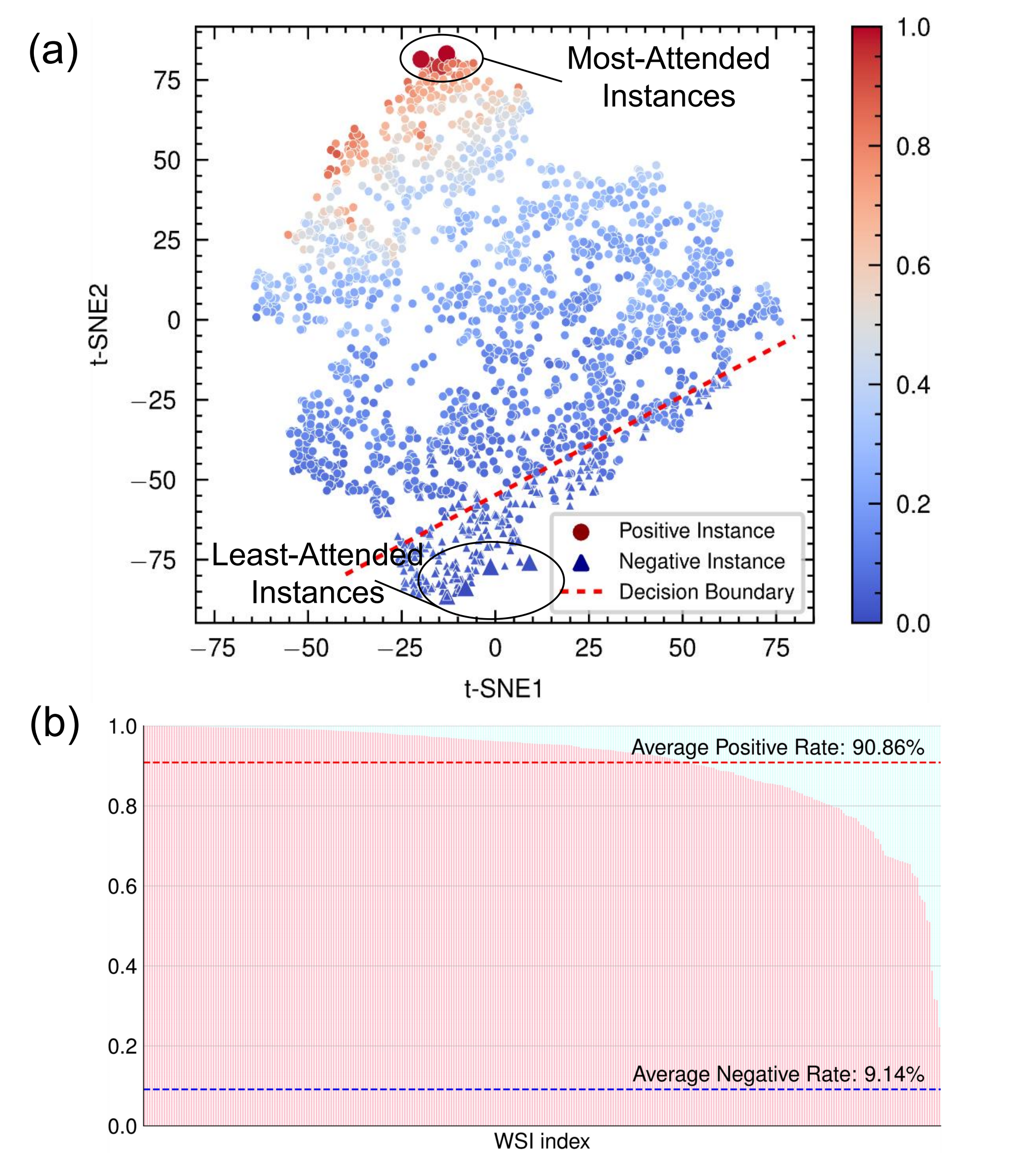}
    \caption{Instance distribution and label proportion visualizations for 1p/19q codeletion. (a) The instance distribution map of a 1p/19q codeletion case. The color represents the attention score, and the shape stands for the predicted label by the OCC strategy. (b) The proportion of positive and negative instances classified by the OCC strategy in all positive WSIs of the TCGA-LGG dataset. Each bar stands for a WSI, where red (blue) indicates the positive (negative) instance proportion.}
    \label{fig:7}
\end{figure}
To study the impact of 1p/19q codeletion on the global morphology, we uniformly select patches from the codeletion WSI and plot the patch features using t-SNE, as shown in Fig. \ref{fig:7}(a). The visualization demonstrates that almost all negative and positive instances are separated by the decision boundary. The most attended and least attended patches are located on the two corners of the plot, and instances with the same distances to the decision boundary have similar attention scores. These findings prove that the OCC strategy can effectively differentiate the negative and positive instances in the 1p/19q codeletion WSI, even without patch-level labels. In addition, in the t-SNE plot, there are several instances below the decision boundary which are classified as negative, i.e., 1p/19q non-codeletion. This phenomenon gives clues that a patient with 1p/19q codeletion might have non-codeletion tissues, which indicates that the codeletion of 1p/19q is in a gradual process rather than affects the whole tissue morphology at once. We further calculate the proportion of positive and negative instances in all 1p/19q codeletion WSIs. The quantitative result shown in Fig. \ref{fig:7}(b) demonstrates that the 1p/19q codeletion instances take up to 90.86\% of areas in a positive WSI on average. These findings further validate that the impact of 1p/19q codeletion has a localized characteristic.

\section{Discussion and Conclusion}
\label{conclusion}
In this paper, we propose a deep learning pipeline dubbed Multi-Beholder to predict the status of five LGG biomarkers with H\&E-stained WSIs. Considering that the instances in negative WSIs are negative, we utilize an OCC strategy to fully exploit these true negative instances to obtain a separable positive-negative instance decision boundary for positive WSIs. Therefore, instances in positive bags can be assigned with accurate instance-level pseudo-labels. Eventually, the instance-level pseudo-labels are utilized to refine the instance features, hence the bag feature aggregated by the refined instance features can be more discriminative. 

Experiments on internal TCGA-LGG and external Xiangya cohort prove the effectiveness of Multi-Beholder. Sensitivity analysis (\textbf{Supplementary Section V}) shows stable performance across random splits and hyperparameters, while leave-one-center-out experiments (\textbf{Supplementary Section VI}) and external validation further demonstrate robustness to domain shift.
Calibration analysis (\textbf{Supplementary Section III}) indicates that, after simple temperature-based post-hoc scaling, our predicted probabilities are well-calibrated and can be interpreted as meaningful confidence estimates for biomarkers.

In addition, DCA in \textbf{Supplementary Section IV} further demonstrates that the Multi-Beholder framework can help in early biomarker triage under a high-sensitivity threshold. Specifically, by analyzing routine digital H\&E slides with our framework, clinicians could obtain predicted risk scores for each biomarker. These scores could be used to prioritize confirmatory molecular tests. Specifically, biomarkers with high predicted risk would be tested earlier, while those with lower risk could be tested later or even omitted, thereby reducing diagnostic costs and expediting diagnosis. Moreover, our model generates attention maps highlighting regions of highest suspicion for each biomarker. These regions can be reviewed by pathologists, aiding interpretation of the model’s predictions and potentially revealing novel morphological patterns linked to the biomarkers.

There are still some limitations to the proposed framework. Our validation is currently based on only a few centers (TCGA and Xiangya), which may constrain generalizability. Future work will focus on multi-center, multi-scanner validation to better characterize real-world robustness and clinical utility. Besides, we observe biomarker-specific performance differences (e.g., MGMT vs. IDH1/2), likely driven by varying label distributions, the unequal strength and consistency of H\&E morphological evidence across biomarkers, and differential sensitivity to inter-center domain shift. Future work will also focus on improving robustness across biomarkers.

\section*{Acknowledgments}
This work was supported in part by the National Natural Science Foundation of China (62031023\&62331011), in part by the Shenzhen Science and Technology Project (GXWD20220818170353009), and in part by the Fundamental Research Funds for the Central Universities (Grant No. HIT.OCEF.2023050). 

\bibliographystyle{ieeetr}
\bibliography{refs.bib}

\begin{IEEEbiography}[{\includegraphics[width=1in,height=1.25in,clip,keepaspectratio]{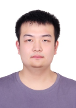}}]{Zijie Fang}
received the BS degree in Software Engineering from Nanjing University of Information Science and Technology in 2022. He is currently working towards the MS degree at Tsinghua Shenzhen International Graduate School, Tsinghua University. His research interests include computational pathology and medical image segmentation.
\end{IEEEbiography}

\begin{IEEEbiography}[{\includegraphics[width=1in,height=1.25in,clip,keepaspectratio]{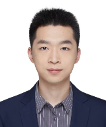}}]{Yihan Liu}
received the MBBS degree from Central South University in 2022. He is currently a Ph.D. candidate in Cancer Biology program at University of Michigan. His research focuses on exploring the role of chromatin modulators and epigenetic dysregulation in cancer progression using genetic and pharmacological tools and multi-omics.
\end{IEEEbiography}

\begin{IEEEbiography}[{\includegraphics[width=1in,height=1.25in,clip,keepaspectratio]{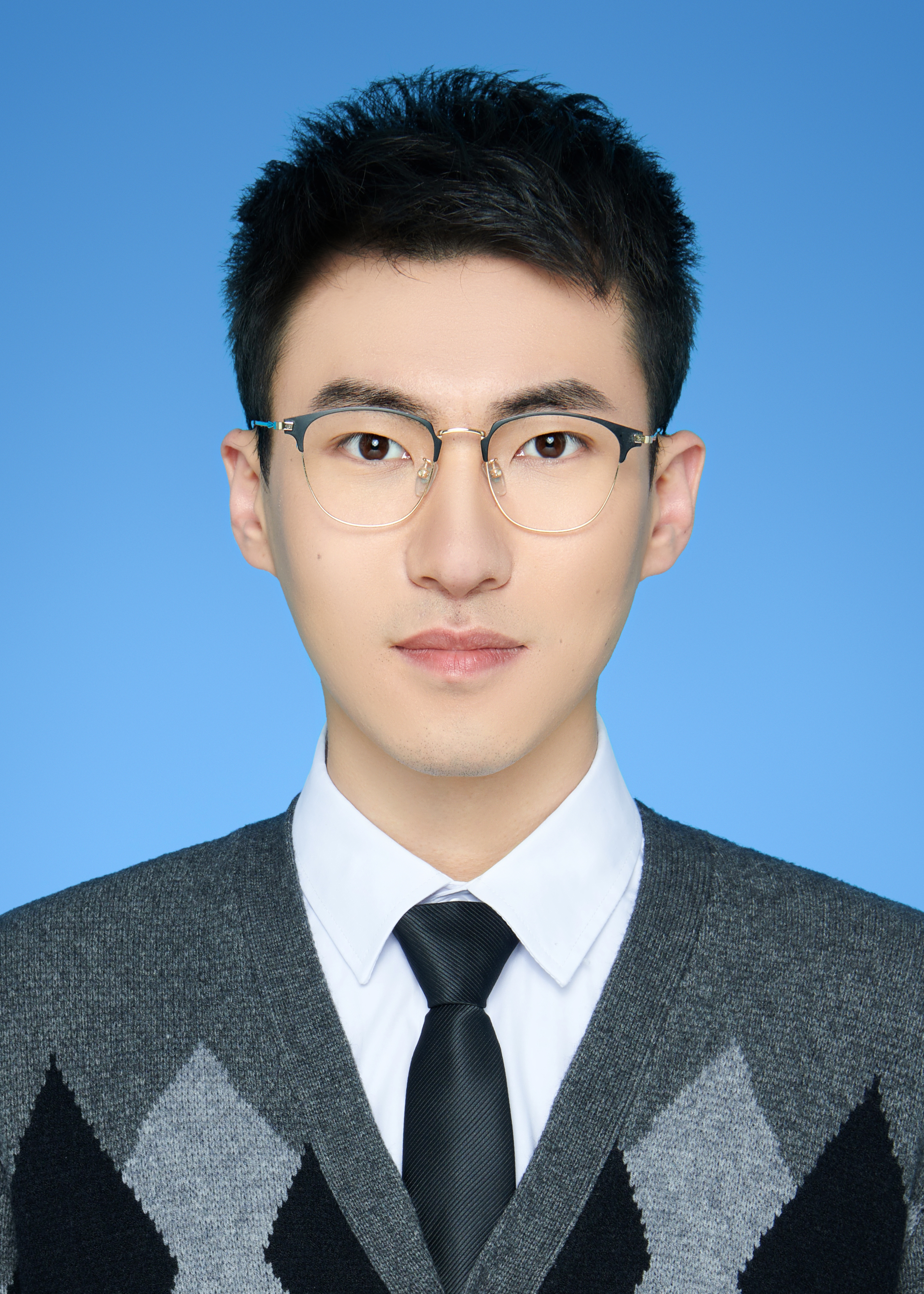}}]{Yifeng Wang}
received the M.S. degree in Harbin Institute of Technology, Shenzhen, China, where he is working toward the Ph.D degree. He is currently pursuing a joint Ph.D. program at the National University of Singapore, funded by the China Scholarship Council. His research interests include deep learning model design and analysis, sensor signal processing, representational learning, and generative deep learning architecture.
\end{IEEEbiography}

\begin{IEEEbiography}[{\includegraphics[width=1in,height=1.25in,clip,keepaspectratio]{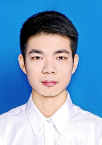}}]{Xiangyang Zhang}
received the MBBS degree from Central South University in 2020, and Master of Clinical Medicine from Central South University in 2023. He is currently a joint Ph.D. student in both Central South University and City University of Hongkong. His research focuses on cancer multi-omics data analysis, medical big data mining, and the relationship between tumor metabolism and immune microenvironment.
\end{IEEEbiography}

\begin{IEEEbiography}[{\includegraphics[width=1in,height=1.25in,clip,keepaspectratio]{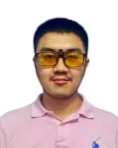}}]{Yang Chen}
received his BS degree in Computer Science and Technology from Ocean University of China in 2021. He is currently pursuing a PhD degree at Tsinghua Shenzhen International Graduate School, Tsinghua University. His research interests include computational pathology and large language models.
\end{IEEEbiography}

\begin{IEEEbiography}[{\includegraphics[width=1in,height=1.25in,clip,keepaspectratio]{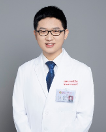}}]{Changjing Cai}
received the MBBS degree and Ph.D. from Central South University. He is currently a clinical doctor and associate professor at Central South University. His research focuses on basic and clinical translational research on accurate diagnosis and treatment of gastrointestinal tumors.
\end{IEEEbiography}

\begin{IEEEbiography}[{\includegraphics[width=1in,height=1.25in,clip,keepaspectratio]{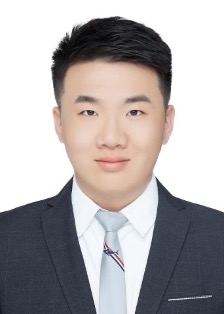}}]{Yiyang Lin}
received his BS degree from Beihang University in 2020 and his MEng degree from Tsinghua University in 2023. He is currently working towards a PhD degree at the Chinese University of Hong Kong, and his research interests include medical image analysis and image generation.
\end{IEEEbiography}

\begin{IEEEbiography}[{\includegraphics[width=1in,height=1.25in,clip,keepaspectratio]{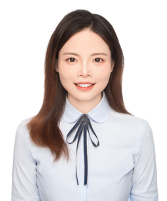}}]{Ying Han}
received the MBBS degree and Ph.D. from Central South University. She is currently a clinical doctor and associate professor at Central South University. Her research focuses on basic and clinical translational research on accurate diagnosis and treatment of gastrointestinal tumors.
\end{IEEEbiography}

\begin{IEEEbiography}[{\includegraphics[width=1in,height=1.25in,clip,keepaspectratio]{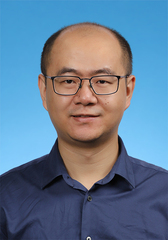}}]{Zhi Wang}
 (S'10-M'14-SM'22) is currently an associate professor at Shenzhen International Graduate School, Tsinghua University. He received his Ph.D. in 2014 and his B.E. in 2008, both from Tsinghua University. His research areas include multimedia networks, mobile cloud computing, and large-scale machine learning systems. He was a recipient of the Natural Science Award of the Ministry of Education (First Prize) in 2017, the National Natural Science Award (Second Prize) in 2018, the Shenzhen Youth Science and Technology Award in 2019, and the Technology Invention Award of the Chinese Institute of Electronics (First Prize) in 2020. In addition, his research won the Best Paper Award of ACM Multimedia, the Best Paper Award of IEEE Transactions on Multimedia, the Outstanding Doctoral Thesis Award of China Computer Federation, the Best Student Paper Award of MMM, and the Best Paper Award of ACM Multimedia, HUMA Workshop. He is an Associate Editor of IEEE TMM and Guest Editor of ACM TIST and JCST. His research has been covered by prestigious technology media, including MIT Technology Review and Synced Review.
\end{IEEEbiography}

\begin{IEEEbiography}[{\includegraphics[width=1in,height=1.25in,clip,keepaspectratio]{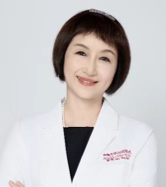}}]{Shan Zeng}
received the MBBS degree and Ph.D. from Central South University. She is currently a clinical doctor and professor at Central South University. Her research focuses on basic and clinical translational research on accurate diagnosis and treatment of gastrointestinal tumors.
\end{IEEEbiography}

\begin{IEEEbiography}[{\includegraphics[width=1in,height=1.25in,clip,keepaspectratio]{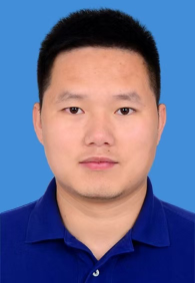}}]{Jun Tan}
received the MBBS degree and Ph.D. from Central South University. He is currently a clinical doctor and assistant professor at Central South University. His research focuses on basic and clinical translational research on accurate diagnosis and treatment of tumors of nervous system.
\end{IEEEbiography}

\begin{IEEEbiography}[{\includegraphics[width=1in,height=1.25in,clip,keepaspectratio]{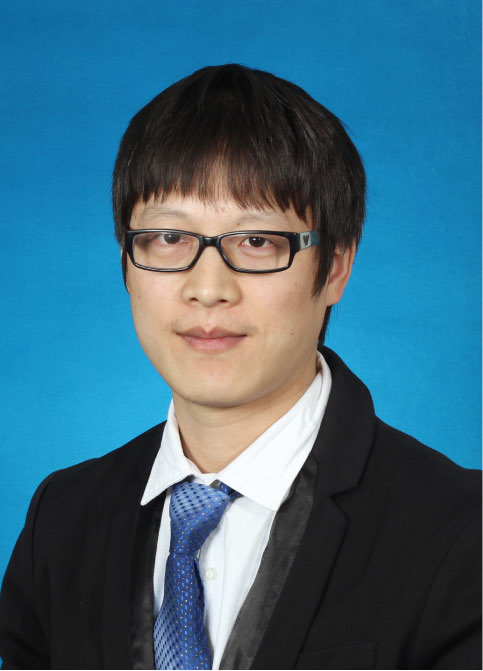}}]{Yongbing Zhang}
(Senior Member, IEEE) received the B.A. degree in English and the M.S. and Ph.D. degrees in computer science from the Harbin Institute of Technology, Harbin, China, in 2004, 2006, and 2010, respectively. From 2010 to 2020, he was with the Tsinghua Shenzhen International Graduate School. From 2016 to 2017, he was a Visiting Scholar with the University of California, Berkeley. He is currently a Professor of computer science and technology with the Harbin Institute of Technology (Shenzhen), Shenzhen, China. His current research interests include signal processing, machine learning, and computational imaging. He was a recipient of the Best Student Paper Award from the IEEE International Conference on Visual Communication and Image Processing in 2015 and the Best Paper Award from the Pacific-Rim Conference on Multimedia in 2018.
\end{IEEEbiography}

\begin{IEEEbiography}[{\includegraphics[width=1in,height=1.25in,clip,keepaspectratio]{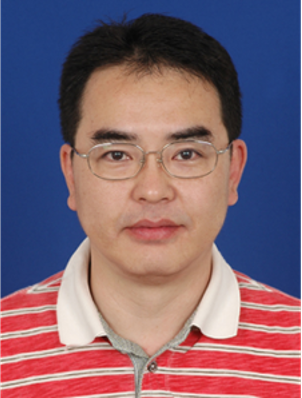}}]{Hong Shen}
received Ph.D. from Central South University. He is currently a professor at Central South University. His research focuses on basic and clinical translational research on accurate diagnosis and treatment of cancers.
\end{IEEEbiography}

\vspace{11pt}

% \bf{If you will not include a photo:}\vspace{-33pt}
% \begin{IEEEbiographynophoto}{John Doe}
% Use $\backslash${\tt{begin\{IEEEbiographynophoto\}}} and the author name as the argument followed by the biography text.
% \end{IEEEbiographynophoto}

% \vfill

\end{document}